\documentclass{amsart}
\RequirePackage{fix-cm}
\usepackage{amsmath}
\usepackage{ulem}
\usepackage{wasysym}
\usepackage{amssymb}
\usepackage{bbold}
\usepackage{appendix}
\usepackage{breakcites,natbib}
\usepackage[margin=1.2in]{geometry}
\usepackage{multirow}
\newcommand{\Earth}{\textrm{E}}

\newcommand{\F}{\textrm{F}}

\newcommand{\norm}[1]{\left\lVert#1\right\rVert}
\newcommand{\ie}{\textit{i.e.,\ }}
\newcommand{\eg}{\textit{e.g.,\ }}
\newcommand{\dd}{\textrm{d}}

\newcommand{\DOF}{DoF\,}

\newcommand{\gal}{\textrm{Gal.}}
\newcommand{\NHIM}{\textrm{NHIM}}
\newcommand{\R}{\textrm{R}}
\newcommand{\Circ}{\textrm{circ}}
\newcommand{\crit}{\textrm{crit}}
\newcommand{\CM}{\textrm{CM}}
\newcommand{\JFAVE}{\langle J_{\textrm{F}} \rangle}
\newcommand{\C}{\textrm{C}}

\newcommand\eval[1]{\begin{array}[t]{@{}c@{\,}|@{\,}}% 
		\raisebox{0pt}[0.85\height][1.33\depth]{$ \displaystyle#1 $}\end{array}}

\newtheorem{rmk}{Remark}
\newtheorem{defi}{Definition}

\usepackage{rotating}

\usepackage[colorlinks,citecolor=blue]{hyperref}

\begin{document}

\title[A deep dive into the $2g+h$ resonance]{A deep dive into the $2g+h$ resonance: 
	separatrices, manifolds and phase space structure of navigation satellites}

\author{J\'er\^ome Daquin         \and
	Edoardo Legnaro \and
	Ioannis Gkolias \and
	Christos Efthymiopoulos 
}

%\authorrunning{J\'er\^ome Daquin et al.} % if too long for running head

\address{
	Department of Mathematics (naXys), 61  Rue de
	Bruxelles, $5000$, Namur, Belgium}
	\email{jerome.daquin@unamur.be}           

 \address{  
	Department of Physics, Aristotle University of Thessaloniki,  $54124$, Thessaloniki, Greece\\
	Research Center for Astronomy and Applied Mathematics, Academy of Athens, Athens,
	$11527$, Greece}

\address{ 
	Department of Physics, Aristotle University of Thessaloniki,  $54124$, Thessaloniki, Greece} 

\address{
	Department of Mathematics, University of Padova, Via Trieste $63$, 35121 Padova, Italy}

\date{\today}

%\maketitle
%\tableofcontents
%\subjclass[2010]{Primary }
%\keywords{medium Earth orbits - numerical integration - keyword 3}
\date{\today}

\maketitle
	
\begin{abstract}  
	Despite extended past studies, several questions regarding the resonant structure of the medium-Earth orbit (MEO) region remain hitherto unanswered. This work describes in depth the effects of the $2g+h$ lunisolar resonance. In particular, (i) 	
	we compute the correct forms of the separatrices of the resonance in the inclination-eccentricity $(i,e)$ space for fixed semi-major axis $a$. This allows to compute the change in the width of the $2g+h$ resonance as the altitude increases. 
	(ii) We discuss the crucial role played by the  value of the inclination of the {\it Laplace plane}, $i_{L}$. Since $i_L$ is comparable to the resonance's separatrix width, the parametrization of all resonance bifurcations has to be done in terms of the {\it proper} inclination $i_{p}$, instead of the mean one. 
	(iii) The subset of circular orbits constitutes an invariant subspace embedded in the full phase space,  the center manifold $\mathcal{C}$, where actual navigation satellites lie. Using $i_p$ as a label, we  compute its range of values for which $\mathcal{C}$ becomes a normally hyperbolic invariant manifold (NHIM). 
	The structure of invariant tori in $\mathcal{C}$ allows 
	to explain the role of the initial phase $h$ noticed in   several works. 
	(iv) Through Fast Lyapunov Indicator (FLI)  cartography, we portray the stable and unstable manifolds of the NHIM as the altitude increases. Manifold oscillations dominate in phase space between $a=24,000$ km and $a=30,000$ km as a result of the sweeping of the $2g+h$ resonance by the $h-\Omega_{{\leftmoon}}$ and $2h-\Omega_{{\leftmoon}}$ resonances. 
	The noticeable effects of the latter are explained as a consequence of  the  relative inclination of the Moon's orbit with respect to the ecliptic. 
	The role of the phases $(h,\Omega_{{\leftmoon}})$ in the structures observed in the FLI maps is also clarified.  	
	Finally,
	(v) we discuss how the understanding of the manifold dynamics could inspire end-of-life disposal strategies.
\end{abstract}

%=============================
\section{Introduction}
%=============================
Twenty years ago S\l{}awomir Breiter in his `concise synthesis'  \citep{sBr01-2} wrote  that 
%---------------------------------------------------------
\textit{`the problem of lunisolar resonances has remained seriously underrated. Quite possibly it was due to the lack of interest in the effects whose typical time scale may exceed decades or even centuries, but this situation have changed once we realised the dangers imposed by space debris'}.  
%---------------------------------------------------------
As prophesied, the situation has changed, the change having been partly boosted by the  questions of how to  `cleverly exploit resonant effects'. This possible exploitation relies  on using natural mechanisms,  pumping-up the eccentricity, guiding satellites in orbital regime where friction could start their decay 
\citep{cCh04,aRo08,lSt15,aRo15,aRo17,rAr18}. This resurgence of interest 
has undoubtedly benefited also from the advances in computing possibilities and the introduction of suited and compact averaged formulations \citep{mLa14,iGk16,mLa21}. These naturally led to
numerical insights for the secular evolution
given the large volume of orbits that can be propagated over $10^{6}+$ orbital revolutions in a few seconds only. Nevertheless, most of the recent endeavours on the topic have departed from the essence of Breiter's analysis \ie from the \textit{systematic} classification of  resonances according to the `fundamental model they match'. The present paper retakes and updates this route, focusing on a case of particular interest for navigation satellites, namely the secular lunisolar inclination-dependent only $2g+h$ resonance. \\

The paper is outlined as follows: 
\begin{enumerate}
	\item Section \ref{sec:model} presents 
	the Hamiltonian model including Earth's $J_2$ and the lunar and solar quadrupolar contributions, averaged in closed form over all short periods. This well-established model is known to encapsulate the complexity of the dynamics at MEOs \citep{jDa16}. 
	The expansion of the Hamiltonian in terms of modified Delaunay action-angle variables is presented, along with a further polynomial expansion in Poincar\'{e} variables around the center of the resonance. Appendix \ref{App:FormalCoeff} gives the most important terms of this expansion, whose detailed form determines all relevant features of dynamics.\\
	
	\item \label{step2}
	Section \ref{sec:IntApproximation} builds upon Breiter's results, encapsulated in the \textit{`Second Fundamental Model'} \citep{sBr01-2}. In this approach, all phase portraits and bifurcations associated with a resonance are parameterized in terms of an approximative resonant integral of motion $I_{\F}$, which, in the case of the $2g+h$ resonance, corresponds to $I_{\F}=G-2H$. Different constant values of $I_{\F}$ define a foliation of surfaces in the action space $(i,e)$. This foliation is exploited to correctly depict in this space the separatrices associated with all possible bifurcations taking place within the $2g+h$ resonance. This, in turn, allows to replace the methodology by which the heuristic drawing of the separatrices was done by \cite{jDa16} with an accurate one. The latter serves to correctly measure the resonance width as a function of the altitude, and, hence, to correct all estimates on the overlapping of the $2g+h$ resonance with other nearby resonances. \\
	
	\item
	Section \ref{sec:NHIM} shows how the analysis of Section \ref{sec:IntApproximation} is perplexed by the fact that the Moon's and Sun's orbits are substantially inclined with respect to the equatorial plane.	
	The existence  of all resonances including the lunar node $\Omega_{\leftmoon}$ is imputable only to the inclination of the Moon's orbit with respect to the ecliptic by $i_{\leftmoon}=5.15^\circ$. This 
	introduces a severe change with respect to theoretical models in which the Moon is considered on the ecliptic (by setting $i_{\leftmoon}=0$), or with a lunar node $\Omega_{\leftmoon}$ frozen to a fixed value (as in \citep{aCe16siam}).
	The effect of the Moon's nodal precession on dynamics is enhanced close to the crossings of the 
	$2g+h$ resonance with the $h-\Omega_{\leftmoon}$ and
	$h-2\Omega_{{\leftmoon}}$ resonances\footnote{In the subsequent we use the shortcut $\mathcal{R}_{k\cdot\sigma}$ to 
		refer to the $k\cdot \sigma$ resonance, \ie the commensurability $k\cdot\dot{\sigma}=0$. The
		$2g+h$, $h-\Omega_{\leftmoon}$ and
		$h-2\Omega_{{\leftmoon}}$ resonances are thus respectively noted $\mathcal{R}_{2g+h}, \mathcal{R}_{h-\Omega_{\leftmoon}}$ and $\mathcal{R}_{h-2\Omega_{{\leftmoon}}}$.}.
	In fact, the corresponding resonant terms in the Hamiltonian are $\mathcal{O}(i_{\leftmoon})$, while the contribution of $i_{\leftmoon}$ in the secular Hamiltonian is $\mathcal{O}(i_{\leftmoon}^2)$ only. 
	
	Properly accounting for the above effects requires exploiting the information contained in the canonical transformation which eliminates all terms non-relevant to the resonance; instead, the `scissor' averaging leading to the integrable model of the resonance used in past studies erases this information. Using the information contained in the canonical transformation leads to understand that $I_{\F}=G-2H$ is not the correct parameter in terms of which the analysis of the resonance phase portraits is to be done. Instead, one has first to quantify the long term evolution of the satellite's inclination vector around its {\it forced equilibrium} value, \ie the inclination $i_{L}$ of the {\it Laplace plane} \citep{vKu97,sTr09}. To this end, we first focus on the ensemble of all possible orbits with $e=0$, which, according to the secular Hamiltonian, are embedded in an invariant subspace of the whole phase space, hereafter called the \textit{center manifold} $\mathcal{C}$ of the circular orbits. We find the following to be a useful description of the invariant dynamics in $\mathcal{C}$: \\
	
	\begin{enumerate}
		\item
		Far from the intersection of the $2g+h$ resonance with any other lunisolar resonance, the dynamics in $\mathcal{C}$ is characterized by a two degrees-of-freedom (\DOF) near-integrable model,  filled with rotational 2D invariant tori. These tori reflect the product (rotation of the satellite's inclination vector around its forced equilibrium) $\times$ (precession of the lunar node). 
		%The associated frequencies, $\dot{h}$ and $\dot{\Omega_{{\leftmoon}}}$ are low-order incommensurable except near $a=24,000$ km and $a=30,000$ km, where, respectively, $\dot{h}-\dot{\Omega_{{\leftmoon}}}\simeq 0$, and $\dot{h}-2\dot{\Omega_{{\leftmoon}}}\simeq 0$. 
		One finds that the evolution of the satellite's inclination vector decouples nearly completely from the motion of the lunar node, so that the satellite's orbital parameters $i(t)$ and $h(t)$ become periodic functions of time, with the frequency $\dot{h}$ only. Then, $h(t)\sim\dot{h}t$ while it can be demonstrated (confer Section \ref{sec:NHIM}) that $i(t)$ takes the form 
		\begin{align}
		i(t)\sim i_p + 
		i_L\cos h(t) +
		c \cos 2h(t), \, c = \mathcal{O}(i_{L}^{2}).
		\end{align}
		Since $i_L$ is a fixed quantity depending only on the model's parameters and the (fixed) $a$, the tori can be parameterized in terms of the quantity $i_p$,  \ie the satellite's {\it proper inclination}. In our analysis we use, equivalently, the Delaunay action $\langle J_F \rangle$.\\
		
		\item
		Close to the intersection of $\mathcal{R}_{2g+h}$ with $\mathcal{R}_{h-\Omega_{\leftmoon}}$ and
		$\mathcal{R}_{h-2\Omega_{\leftmoon}}$ (near $a=24,000$ km and $a=30,000$ km respectively), the dynamics in $\mathcal{C}$ is characterized also by a  2-\DOF \, near-integrable model  filled both with rotational and librational tori. This requires a different parametrization for the 2D tori in $\mathcal{C}$, taking into account both angles $h$ and $\Omega_{{\leftmoon}}$. It will be demonstrated that it is, precisely, the effect of these resonances which explains the dependence of the $2g+h$ FLI cartography not only on the initial phase $h$, but also on the initial phase $\Omega_{{\leftmoon}}$ as observed numerically  \citep{eAl16,aRo17}.\\ 
	\end{enumerate}
	
	\item
	In sections \ref{sub:NHIM}  and \ref{sec:NHIM-Escapers}, we address the existence of 
	large chaotic seas for
	$a>24,000$. 
	\cite{jDa16} and \cite{aCe16} advocated  Chirikov's heuristic overlap of nearby resonances \citep{bCh79} as the generic mechanism for the appearance of macroscopic chaos in MEOs. Here, instead, we clarify the role of particular invariant hyperbolic structures in phase space 
	leading to instability for  the specific  $2g+h$ resonance. The cornerstone in this process is the identification of the intricate structure of the stable and unstable invariant manifolds in phase space emanating from a subset $\mathcal{C}_{\NHIM}$ of the center manifold $\mathcal{C}$ being \textit{normally hyperbolic}, as  pointed out by \cite{aCe16siam}. 
	Our key result regarding the NHIM and its stable and unstable manifolds is that the parameterization of the tori of $\mathcal{C}$ (in particular close to the double resonance crossings with 
	$\mathcal{R}_{h-\Omega_{{\leftmoon}}}$ and 
	$\mathcal{R}_{h-2\Omega_{{\leftmoon}}}$), allows to theoretically predict the domain of initial conditions, in the space $(i,h,\Omega_{{\leftmoon}})$ for $e=0$, which belongs to the normally hyperbolic set $\mathcal{C}_{\NHIM}$. 
	In fact, for altitudes higher than $a=24,000$ km, this prediction can be used, \textit{per se}, for the identification of the width in inclination $\Delta i_R$ of the resonance. Besides interpreting the dependence of the location and width of the resonance on the initial phases $h,\Omega_{{\leftmoon}}$, these results show that, for $a>24,000$ km, the stable and unstable manifolds of $\mathcal{C}_{\NHIM}$ are developed structures which dominate the phase space, hence, rendering inapplicable predictions based on the separatrices of Breiter's `integrable' model of resonance. Furthermore, the fractal distribution of the initial conditions leading to escape is characterised.
\end{enumerate}
Before delving further into the details, let us underline that whilst  $\mathcal{R}_{2g+h}$ is one of the most important for MEO dynamics, the analysis presented here applies generically to inclination-dependent only resonances. In particular, the condition $e=0$ defines an invariant centre manifold for all values of $(i,h,\Omega_{{\leftmoon}})$. Revealing how particular secondary resonances involving the lunar node affect the centre-manifold dynamics requires a case by case treatment proposed for further study.

%-----------------
\section{The Hamiltonian model}\label{sec:model}
%-----------------
The perturbations acting on the test-particle include the  secular contribution of the main zonal Earth's disturbance (the $J_{2}$ effect) and external disturbances due to the Moon and the Sun, both truncated to the order $l=2$ in their Legendre's polynomial expansion (quadrupolar assumption). The model uses the Earth equatorial plane as a reference frame. The final secular model depends explicitly on the time through the ecliptic lunar node $\Omega_{{\leftmoon}}$ whose time variation is assumed to be linear with time, $\Omega_{{\leftmoon}}(t)=n_{\Omega_{{\leftmoon}}}t + \Omega_{{\leftmoon}}(0)$, with $n_{\Omega_{{\leftmoon}}} <0$. The ecliptic lunar nodal period $T_{\Omega_{\leftmoon}}=2\pi/\vert n_{\Omega_{{\leftmoon}}}\vert$ is taken equal to $18.6$ years. The values $i_{{\leftmoon}}=5^{\circ}15$ for the Moon's inclination with respect to the ecliptic plane and  $\varepsilon=23^{\circ}44$ for the the obliquity of the ecliptic are adopted.\\

The secular model is derived from the full Hamiltonian 
\begin{align}\label{eq:H}
\mathcal{H}=
\mathcal{H}_{\textrm{Kep}} 
+ 
\mathcal{H}_{J_{2}}
+
\mathcal{H}_{\odot}
+
\mathcal{H}_{{\leftmoon}}, 
\end{align} 
where
\begin{align}
&	\mathcal{H}_{\textrm{Kep}} = \frac{v^{2}}{2}-\frac{\mu_{\Earth}}{r}, \label{eq:Hkep}\\
&	\mathcal{H}_{J_{2}} = \frac{J_{2}\mu_{\Earth}R_{\Earth}^{2}(3\sin^{2}\phi -1)}{2r^{3}},  \\
& \mathcal{H}_{\odot} = 
-\frac{\mu_{\odot}}{r_{\odot}}
\Big(
\frac{r_{\odot}}{\norm{\bold{r}-\bold{r}_{\odot}}}
-
\frac{\bold{r} \cdot \bold{r}_{\odot}}{r_{\odot}^{2}}
\Big), \\
& \mathcal{H}_{{\leftmoon}} = 
-\frac{\mu_{{\leftmoon}}}{r_{\leftmoon}}
\Big(
\frac{r_{{\leftmoon}}}{\norm{\bold{r}-\bold{r}_{{\leftmoon}}}}
-
\frac{\bold{r} \cdot \bold{r}_{{\leftmoon}}}{r_{{\leftmoon}}^{2}}
\Big).\label{eq:HMoon}
\end{align}
In Eqs.\,(\ref{eq:Hkep}-\ref{eq:HMoon}), $r,v$ denotes the geocentric distance and velocity of the test-particle, $\phi$ is the  geocentric latitude and $r_{\Earth}$ is the mean equatorial Earth's radius, $J_{2}$ the Earth's oblateness coefficient. The geocentric distances of the Moon and Sun are respectively noted $r_{{\leftmoon}}, r_{\odot}$, their geocentric vectors are $\bold{r}_{{\leftmoon}}, \bold{r}_{\odot}$. The gravitational parameters are noted  $\mu_{\Earth}, \mu_{{\leftmoon}}, \mu_{\odot}$. Finally, $\norm{\bullet}$ denotes the euclidian norm. The physical values of the coefficients used in this study are provided in  Appendix \ref{App:FormalCoeff}. 

%---------------------------------------
\subsection{Secular Hamiltonian}
%---------------------------------------
Denoting by $(a,e,i,\ell,\omega,\Omega)$ the Keplerian elements (semi-major axis, eccentricity, inclination, mean anomaly, argument of the perigee, right ascension of the ascending node), a secular Hamiltonian model can be arrived at by averaging Eq.\,(\ref{eq:H}) over the satellite's Moon's and Sun's mean anomalies $\ell, \ell_{\odot}, \ell_{{\leftmoon}}$ (fast variables). The model is given in closed form over eccentricities using ad hoc differential relationships involving the true anomalies $f,f_{{\leftmoon}},f_{\odot}$ and the eccentric anomaly $E$ \citep{wKa66}. We have
\begin{align}\label{eq:final_ham}
\mathcal{\bar{\bar{H}}}= \mathcal{H}_{\textrm{Kep}}+
\bar{\mathcal{H}}_{J_{2}}
+
\bar{\bar{\mathcal{H}}}_{\odot}
+
\bar{\bar{\mathcal{H}}}_{{\leftmoon}},
\end{align}
where $\mathcal{H}_{\textrm{Kep}}=-\mu_{\Earth}/(2a)$ and  
\begin{align}
& \bar{\mathcal{H}}_{J_{2}}=
\frac{1}{2\pi}\int_{0}^{2\pi}
\mathcal{H}_{J_{2}} \, \dd \ell =\frac{1}{2\pi}\int_{0}^{2\pi}\mathcal{H}_{J_{2}} \frac{r^2}{a^2\sqrt{1-e^2}} \, \dd f, \\
& \bar{\mathcal{H}}_{\odot}=
\frac{1}{2\pi}\int_{0}^{2\pi}\mathcal{H}_{\odot}  \, \dd \ell =
\frac{1}{2\pi}\int_{0}^{2\pi}\mathcal{H}_{\odot} (1-e\cos E) \, \dd E, \\
& \bar{\bar{\mathcal{H}}}_{\odot}=\frac{1}{2\pi}\int_{0}^{2\pi}
\bar{\mathcal{H}}_{\odot} \, \dd \ell_{\odot}
=
\frac{1}{2\pi}\int_{0}^{2\pi}
\bar{\mathcal{H}}_{\odot} \frac{r_{\odot}^2}{a_{\odot}^2\sqrt{1-e_{\odot}^2}} \, \dd f_{\odot},\\
& \bar{\mathcal{H}}_{{\leftmoon}}=
\frac{1}{2\pi}\int_{0}^{2\pi}\mathcal{H}_{{\leftmoon}}  \, \dd \ell =
\frac{1}{2\pi}\int_{0}^{2\pi}\mathcal{H}_{{\leftmoon}} (1-e\cos E) \, \dd E, \\
& \bar{\bar{\mathcal{H}}}_{{\leftmoon}}=\frac{1}{2\pi}\int_{0}^{2\pi}
\bar{\mathcal{H}}_{{\leftmoon}} \, \dd \ell_{{\leftmoon}}
=
\frac{1}{2\pi}\int_{0}^{2\pi}
\bar{\mathcal{H}}_{{\leftmoon}} \frac{r_{{\leftmoon}}^2}{a_{{\leftmoon}}^2\sqrt{1-e_{{\leftmoon}}^2}} \, \dd f_{{\leftmoon}}.
\end{align}

The Hamiltonian $ \bar{\bar{\mathcal{H}}}$ is expressed in terms of the Delaunay action-angle variables
\begin{align}
\left\{
\begin{aligned}
& L=\sqrt{\mu \, a}, \hspace{1.68cm} \ell=M, \\
& G=L \, \sqrt{1-e^{2}}, \hspace{0.77cm} g=\omega,\\
& H= G \, \cos i, \hspace{1.24cm} h=\Omega,
\end{aligned}
\right.
\end{align}
or the modified Delaunay action-angle variables 
\begin{align}
\left\{
\begin{aligned}
& \Lambda=L, \hspace{1.65cm} \lambda= \ell + \varpi, \\
& P=L-G, \hspace{0.94cm} p=-\varpi,\\
& Q=G-H, \hspace{0.89cm} q=-h,
\end{aligned}
\right.
\end{align}
where $\varpi=g+h$. The use of action-angle variables presents several benefits stemming from the analytical aparatuses associated with the canonical formalism. Note that for small eccentricities and inclinations, $P = \mathcal{O}(e^2)$ and $Q = \mathcal{O}(i^2)$. Due to the secular invariance of the Delaunay momentum $L$ ($\lambda$ is ignorable), or equivalently the semi-major axis $a$, the final dynamics  has 2-\DOF and depends explicitly on the time $t$ through the longitude of the lunar node $\Omega_{{\leftmoon}}$ with respect to a fixed axis on the ecliptic. Thus the Hamiltonian $\bar{\bar{\mathcal{H}}}_{\textrm{sec}}$ is of 2.5-\DOF. The phase space is extended by introducing a dummy action variable $J_{{\leftmoon}}$ conjugate to the angle $\Omega_{{\leftmoon}}$. Then, the set of variables $(P,Q,J_{{\leftmoon}},p,q,\Omega_{{\leftmoon}})$ defines an autonomous  3-\DOF Hamiltonian
\begin{align}\label{eq:FinalHsec}
\mathcal{H}_{\textrm{sec}}(P,Q,J_{\leftmoon},p,q,\Omega_{{\leftmoon}};a(L)) = \bar{\bar{\mathcal{H}}}_{\textrm{sec}}(P,Q,p,q,\Omega_{{\leftmoon}};a(L)) + n_{\Omega_{{\leftmoon}}} J_{\leftmoon}.
\end{align}  
Finally, we denote by $\mathcal{H}_{0}$ the sum of $\mathcal{H}_{\textrm{Kep}}$ and of the action-dependent only part of $\mathcal{H}_{\textrm{sec}}$ and decompose formally the Hamiltonian as
\begin{align}\label{eq:FinalH}
\mathcal{H}(P,Q,J_{\leftmoon},p,q,\Omega_{{\leftmoon}};a(L))=
\mathcal{H}_{0}(P,Q,J_{\leftmoon};a(L)) 
+ 
\mathcal{H}_{1}(P,Q,J_{\leftmoon},p,q,\Omega_{{\leftmoon}};a(L)).
\end{align}
One has 
\begin{equation}\label{eq:h0}
\mathcal{H}_0 = \mathcal{H}_{\textrm{Kep}}+\mathcal{H}_{J2} + n_{\Omega_{{\leftmoon}}} J_{\leftmoon}+ \mathcal{H}_{0,{\leftmoon}} + \mathcal{H}_{0,\odot},
\end{equation}
where 
\begin{equation}\label{eq:hj2}
\mathcal{H}_{J_{2}}(G,H;L)= -{3\over 4}\mu_{\Earth}r_{\Earth}^2J_{2}L^{-3} (H^2-G^2)G^{-5}
\end{equation}
is the (averaged over the fast angle $\ell$) contribution of the Earth's $J_2$ term, 
while $\mathcal{H}_{0,{\leftmoon}} = \mathcal{O}(\mu_{{\leftmoon}}a^2/a_{{\leftmoon}}^3)$ and 
$\mathcal{H}_{0,\odot} = \mathcal{O}\left(\mu_{\odot}a^2/a_{\odot}^3\right)$ are the terms 
of the lunisolar disturbing function free on the canonical angles. 

%-----------------------------
\subsection{Lunisolar resonances}\label{sub:LSessentials}
The Hamiltonian (\ref{eq:FinalH}) contains trigonometric terms with arguments
\begin{align}\label{eq:Com}
\sigma_{k}= k \cdot \upsilon =
k_{1}g + k_{2}h + k_{3} \Omega_{{\leftmoon}}, 
k_{1} \in \{-2,0,2\}, 
(k_{2},k_{3}) \in \{-2,-1,0,1,2\}^{2},
\end{align}
leading to a resonance whenever  
$
\dot{\sigma}_{k} \approx 0.
$
The frequencies $\dot{g}$ and $\dot{h}$ involved in the time derivative of Eq.\,(\ref{eq:Com}) are dominated by the $J_{2}$ effects and read
\begin{align}\label{eq:omedot}
\left\{
\begin{aligned}
& \dot{g}=  \partial_{G} \mathcal{H}_{0}\approx\partial_{G}\mathcal{H}_{J_{2}}=
\frac{3}{4}J_{2}n\frac{r_{\Earth}^2}{a^2}
\frac{5 \cos^{2}i-1}{(1-e^2)^2},\\
& \dot{h}= \partial_{H} \mathcal{H}_{0}\partial_{H}\approx\partial_{H} \mathcal{H}_{J_{2}}=
-\frac{3}{2}J_{2}n\frac{r_{\Earth}^2}{a^2}
\frac{\cos i}{(1-e^2)^2}.
\end{aligned}
\right.
\end{align} 
For a fixed wavenumber vector $k$, the resonance condition $\mathcal{R}_{\sigma_{k}}$ takes the form \begin{align}
f_{k}(a(L),e(L,P),i(L,P,Q))=0,
\end{align}
therefore it defines a 2D-surface in the space of the elements $(a,e,i)$ or in the space of the corresponding actions $(L,P,Q)$. Such a surface is called the `resonant manifold' corresponding to the resonance labelled $k$ and have been portrayed in several instances \citep{aCo62,sHu80,tEl96,sBr01-2}.  In the approximation of Eqs.\,(\ref{eq:omedot}), resonances $\mathcal{R}_{k_1g+k_2h}$ (free of the lunar node) satisfy the relation 
\begin{equation}\label{eq:resinc}
k_1(5\cos^2i-1)-2k_2\cos i = 0,  
\end{equation}
and thus depend only on the inclination \citep{sHu80}. In terms of both their influence to the dynamics, as well as effects on the long-term orbital behavior of navigation satellites, the two most important inclination-dependent resonances at MEO are $\mathcal{R}_{2g+h}$ and $\mathcal{R}_{2g}$.  
%While we provide results for a large part of the MEO domain (from $a=20,000$ km to $a=30,000$ km), we give particular details for $\mathcal{R}_{2g+h}$ in the region around $a = a_{\gal} \sim 4.64\, r_{\Earth} = 29,600$ km where the European navigation constellation Galileo operates.  

%==========================================================
\section{Integrable approximation of the $2g+h$ resonance}
\label{sec:IntApproximation}
%==========================================================
Departing from \cite{sBr01-2}, we now present how we construct an integrable model for $\mathcal{R}_{2g+h}$ allowing to theoretically trace all critical bifurcations around the resonance's critical inclination 
$i_{\star}=56.06^\circ$. Most notably, by taking into account that all resonant motions are stratified along a foliation of so-called \textit{planes of fast drift}, we obtain the correct form of the curves representing the intersection of the separatrices of  $\mathcal{R}_{2g+h}$ with the plane $(i,e)$ for fixed $a$, correcting the heuristic drawing of the separatrices of past works (in particular Fig.\,$2$, $3$ and $13$ of \cite{jDa16}). 
%The method is applicable to any lunisolar resonance up to a redefinition of the corresponding plane of fast drift.  
%-----------------------
\subsection{The resonant Hamiltonian}
%-----------------------
Setting $k_1=2$ and $k_2=1$ in Eq.\,(\ref{eq:resinc}), we recover as root the critical inclination $i=i_{\star}=56.06^\circ$ satisfying
\begin{equation}\label{eq:cosiq}
\cos i_{\star} = {1\over 10}\left(1+\sqrt{21}\right). 
\end{equation}
Setting a fixed reference value $a$ of the semi-major axis, the corresponding Delaunay action for a circular orbit ($e=0$) is $Q_{\star} = \sqrt{\mu_{\Earth}a}(1-\cos i_{\star})$. Treating the quantities $\delta Q=Q-Q_{\star}$ and $P$ as small, the Hamiltonian (\ref{eq:FinalH}) can be developed in powers of $(\delta Q,P)$. The truncated expansion up to order $N$ in the action variables $\delta Q,P$ reads
\begin{eqnarray}\label{eq:hamexp}
\mathcal{H} &= &n_q\delta Q + n_p P + n_{\Omega_{\leftmoon}} J_{{\leftmoon}} + 
\sum_{s=0}^N\sum_{k_2=0}^4\sum_{k_3=-2}^2
\mathcal{P}_{s,0,k_2,k_3}(\delta Q,P)\cos(k_2q+k_3\Omega_{{\leftmoon}}) + \nonumber \\
~&+&P\left(\sum_{s=0}^{N-1}\sum_{k_2=-4}^4\sum_{k_3=-2}^2
\mathcal{P}_{s,2,k_2,k_3}(\delta Q,P)\cos(2p+k_2q+k_3\Omega_{{\leftmoon}})\right),
\end{eqnarray}
where $n_q$, $n_p$ are calculated as $n_q = -\dot{\Omega}$, $n_p = -\dot{\omega}-\dot{\Omega}$, using Eq.\,(\ref{eq:omedot}) with $i=i_\star$, $e=0$, and $\mathcal{P}_{s,0,k_2,k_3}(\delta Q,P)$, 
$\mathcal{P}_{s,2,k_2,k_3}(\delta Q,P)$ are polynomials of degree $s$ in $\delta Q,P$. 
Finally, the resonant combination $2g+h$ corresponds to $2g+h = 2(q-p)-q=q-2p$. Thus, the resonant condition reads $n_q=2n_p$. 

An integrable model for $\mathcal{R}_{2g+h}$ can be obtained via a near-identity canonical transformation eliminating all trigonometric terms in the Hamiltonian (\ref{eq:hamexp}) except those satisfying the conditions $k_2=-1$, $k_3=0$ or $k_2=k_3=0$. Basic perturbation theory (see \eg \cite{cEf11}) guarantees that i) such a transformation exists at formal level, since $n_q=2n_p$, ii) with an abuse of notation, using the same symbols for the variables $(\delta Q,P,q,p)$ before and after one normalization step, the resulting Hamiltonian is derived from the Hamiltonian (\ref{eq:hamexp}) by the `scissor' rule, \ie dropping all terms not satisfying any of the resonant conditions. Then, we arrive at the Hamiltonian model of resonance
\begin{equation}\label{eq:h2gh}
\mathcal{H}_{2g+h} = n_{\Omega_{\leftmoon}} J_{{\leftmoon}}+ \mathcal{H}_{\R},
\end{equation}
where
\begin{equation}\label{eq:hamres21}
\mathcal{H}_{\R} = n_q\delta Q + n_p P + 
\sum_{s=0}^N\mathcal{P}_{s,0,0,0}(\delta Q,P)
+P\left(\sum_{s=0}^{N-1}\mathcal{P}_{s,2,-1,0}(\delta Q,P)\cos(2p-q)\right).
\end{equation}

%------------------------------------
\subsection{Plane of fast drift}
%------------------------------------
Since the angle $\Omega_{{\leftmoon}}$ was eliminated, $J_{{\leftmoon}}$ becomes an irrelevant constant for the dynamics and we thus focus on the dynamics of the 2-\DOF\, Hamiltonian (\ref{eq:hamres21}). The key remark is that the Hamiltonian $\mathcal{H}_{\R}$ is integrable, since the quantity
\begin{equation}\label{eq:jf}
J_{\F} = \delta Q + P/2,  
\end{equation}
is a first integral of motion of the flow under $\mathcal{H}_{\R}$ independent of and in involution with the energy $E=\mathcal{H}_{\R}$. 

Similarly as in the classical Kozai-Lidov mechanism \citep{mlLi62,yKo62}, the existence of a first integral $J_{\F}$ implies that, under the integrable approximation (\ref{eq:hamres21}), all trajectories at the $2g+h$ resonance are subject to coupled variations between the eccentricity and inclination. In fact, the relation 
\begin{equation}\label{eq:jf2}
J_{\F}(a,e,i) = \sqrt{\mu_\Earth a}\left(\sqrt{1-e^2}(\cos i - \cos i_\star) +{1\over 2}(1-\sqrt{1-e^2})
\right) = \textrm{const}
\end{equation}
defines a 2D surface in the space $(a,e,i)$. Since $a$ is constant in the secular dynamics, at every fixed value of $a$ one obtains a foliation of curves in the plane $(i,e)$. The curves are parabola-like since, setting $\Delta i = i-i_\star$,  Eq.\,(\ref{eq:jf2}) implies
\begin{align}\label{eq:jfcurve}
J_{\F}
= 
\sqrt{\mu_\Earth a}\left({e^2\over 2}(1/2-\cos i_{\star})
-
\Delta i \, \sin i_{\star}  +
\mathcal{O}(e^2\Delta i)+\mathcal{O}(\Delta i^2)\right).
\end{align}
The central panel of Fig.\,\ref{fig:fig1} shows several curves of Eq.\,(\ref{eq:jf2}) for $a=20,000$ km in a small interval of values around the critical inclination. One immediately notices the effectiveness of the Kozai-Lidov locking of eccentricity with inclination in causing 
\textit{integrable transport}: a fast rise in eccentricity  takes place along each plane of fast drift.
As an estimate, at the center of the resonance ($J_{\F}=0$), along the plane of fast drift the eccentricity scales with the inclination approximately as 
\begin{align}\label{Eq:EccIncCoupled}
e^2 \approx \frac{2 \Delta i \sin i_{\star}}{\cos i_{\star}-1/2}.
\end{align}  
Rather coincidentally, the denominator in Eq.\,(\ref{Eq:EccIncCoupled}) has a small numerical value equal to $\cos i_\star -1/2 = 0.058$, implying, \eg for $a=a_{\gal}=29,600$ km, that only a variation of $\Delta i \sim 1.2^{\circ}$ suffices for the eccentricity to grow to the re-entry value  $e_{\textrm{re-entry}}=1-R_{\Earth}/a \sim 0.78$
\footnote{One needs to remark, however, that the integrable approximation only serves to obtain quantitative estimates on the allowed maximum size of eccentricity growth inside the resonance. The sensitive dependence of this phenomenon on the initial conditions, discussed in numerical studies is to be interpreted, instead, in an entirely different context, namely the intricate structure of the asymptotic manifolds emanating from the center manifold of the full (non-integrable) Hamiltonian model, see \cite{iGk19} and section \ref{sec:NHIM-Escapers} below.}.

Let us remark that, albeit obtained in the context of integrable dynamics, visual comparison shows that the foliation of the planes of fast drift given by Eq.\,(\ref{eq:jfcurve}) represents well the limits of the resonance in the $(i,e)$ plane even when chaos is present, as revealed using finite time dynamical chaos indicators (see \eg \cite{jDa16}, Fig.\,9). A more precise analysis shows that the limits of the resonance are actually related to the sequence of bifurcations of critical points taking place in phase space as the value of $J_{\F}$ is altered. To this analysis we now turn our attention.

\subsection{Stability of the circular orbit and resonance width}\label{sub:StabilityCirc}
%----------------------------------------------------
We now proceed to a study `\`{a} la Breiter' of the phase portraits associated with the resonance $\mathcal{R}_{2g+h}$. To this end, we derive a polynomial expression for the Hamiltonian $\mathcal{H}_{\R}$ in Poincar\'{e} canonical variables $(X,Y)$, in which $J_{\F}$ is a parameter. Consider, first, the canonical transformation 
$(\delta Q,P,q,p,J_{\leftmoon},\Omega_{\leftmoon})$  $\rightarrow$ $(J_\R,J_\F,J_{\leftmoon},u_\R,u_\F,u_{\leftmoon})$ defined by
\begin{align}
\left\{
\begin{aligned}\label{eq:ResCoordF}
&J_\R=P, \hspace{2cm} u_\R=p-q/2,  \\
&J_\F=P/2 + \Delta Q,\hspace{0.68cm} u_\F=q, \\
& J_{\leftmoon}=J_{\leftmoon},\hspace{1.72cm} u_{\leftmoon} = -\Omega_{\leftmoon}.
\end{aligned}
\right.
\end{align}
The above subscripts $(\R,\F)$ stand for `resonant' and `fast' (variables) respectively, since the resonant angle $u_\R$ undergoes time variations which are slow (since $\dot{u}_\R\approx 0$) with respect to the variations of the angle $u_\F$. We also note that $J_\F$ in (\ref{eq:ResCoordF}) coincides with the definition of the Kozai-Lidov integral (\ref{eq:jf}).  

All terms $\mathcal{P}_{s,2,-1,0}(\delta Q,P)\cos(2p-q)$ in Eq.\,(\ref{eq:hamres21}) take now the form $J_{\R}\cos(2u_{\R})\mathcal{P}_{s,2,-1,0}(J_\R,J_\F)$, 
with the functions $\mathcal{P}_{s,2,-1,0}$ polynomial in both their arguments. Then, defining the Poincar\'{e} canonical variables 
\begin{align}\label{Eq:ResCoord}
\left\{
\begin{aligned}
X=\sqrt{2J_\R}\sin u_\R,\\
Y=\sqrt{2J_\R}\cos u_\R,
\end{aligned}
\right.
\end{align} 
one has $J_\R\cos(2u_\R) = Y^2-X^2$, implying that the Hamiltonian $H_\R$ becomes polynomial in $(X,Y)$. Up to degree 3 in the action variables $(J_\R,J_\F)$ the Hamiltonian reads:
\begin{eqnarray}\label{eq:H1dof}
\mathcal{H}_{\R}&=&b_{10}J_\F +
\sum_{i=1}^{3} c_{(2i)0} X^{2i} 
+
\sum_{i=1}^{3} c_{0(2i)} Y^{2i}
+ 
c_{120}J_\F X^2 + c_{102} J_\F Y^2 +(c_{22}+ c_{122}J_{\F})X^2Y^2 \nonumber\\
&+&b_{20}J_\F^2 + c_{42}X^4Y^2 + c_{24}X^2Y^4+
c_{220}J_{\F}^{2}X^{2} + 
c_{202}J_{\F}^{2}Y^{2} +c_{140}J_{\F}X^{4}+c_{104}J_{\F}Y^{4}~.
\end{eqnarray}
The most important  coefficients entering Eq.\,(\ref{eq:H1dof}) are given in Table \ref{Tab:HR} of Appendix \ref{App:FormalCoeff} as a function of the model's parameters. 

Hamilton's equations for Eq.\,(\ref{eq:H1dof}) yield $\dot{X}=\dot{Y}=0$ for $X=Y=0$. Therefore, $P_0 = (X,Y)=(0,0)$ is a fixed point whatever the value of 
$J_{\F}$. The point $P_0$ corresponds to $J_\R = P = 0$, implying $e=0$. Therefore, it labels a circular orbit at inclination
\begin{align}
i_{\Circ} = \arccos\left(1-\frac{J_{\F}+Q_{\star}}{L}\right).
\end{align}
Actual GPS orbits or Galileo satellites are very close to such circular orbits, \ie to the fixed point of the secular dynamics. 
The linear stability of the circular orbits is determined by the eigenvalues of the Jacobian matrix of the linearization of the Hamiltonian vector field of Eq.\,(\ref{eq:H1dof}) at the origin given by 
\begin{align}\label{eq:JacAna}
\mathcal{J}(0,0)
= 
\begin{pmatrix}
0 					 & 2c_{02}+2c_{102}J_{\F} \\
-2c_{20}-2c_{120}J_{\F}& 0
\end{pmatrix}+\mathcal{O}\left(J_\F^2\right). 
\end{align}
Taking into account that 
\begin{equation}
c_{02}=-c_{20}={15a^{3/2}(1+\cos i_\star)\sin i_\star \sin 2\varepsilon
	\over 32\mu_\Earth^{1/2}}
\left[{\mu_{\leftmoon}\over a_{\leftmoon}^3(1-e_{\leftmoon}^2)^{3/2}}\left(1-{3\over 2}\sin^2 i_{\leftmoon}\right)
+{\mu_{\odot}\over a_{\odot}^3(1-e_{\odot}^2)^{3/2}}\right]
\end{equation}
while $c_{102}$ and $c_{120}$ are equal to leading order $\mathcal{O}(J_2)$ and differ only in terms $\mathcal{O}(\mu_{\leftmoon}/a_{\leftmoon}^3)$, $\mathcal{O}(\mu_{\odot}/a_{\odot}^3)$, 
\begin{equation}
c_{102} = c_{102}+\mathcal{O}\left({\mu_b\over a_b^3}\right),
c_{120} = c_{102}+\mathcal{O}\left({\mu_b\over a_b^3}\right), 
c_{102} = {3J_2\mu_\Earth^{1/2}R_\Earth^2\cos i_\star\over 2a^{7/2}}
\end{equation}
the characteristic polynomial,
\begin{align}\label{eq:PC}
P(\lambda)=\lambda^{2}-(2c_{02}+2c_{102}J_{\F})
(-2c_{20}-2c_{120}J_{\F}),
\end{align}
has two imaginary roots when $J_\F < J_{\F,\min}$ or $J_{\F}>J_{\F,\max}$, while it has two real roots for $J_{{\F},\min<}J_\F<J_{\F,\max}$, where
\begin{equation}\label{eq:jfminmax}
J_{\F,\min} = -c_{20}/c_{102}+\mathcal{O}\left({\mu_b\over a_b^3}\right),
J_{\F,\max} = c_{20}/c_{102}+\mathcal{O}\left({\mu_b\over a_b^3}\right).
\end{equation}
Taking the corresponding inclinations of the circular orbits at $J_\F = J_{\F,\min}$, $J_\F = J_{\F,\max}$, we find the limits in inclination for which the circular orbit is linearly unstable. To leading order in $J_\F$ we arrive at
\begin{equation}\label{eq:iminmax}
i_{\max,\min} = i_\star\pm{5a^5(1+\cos i_\star)\sin 2\varepsilon
	\over
	4(10\cos i_\star-1)\mu_\Earth J_2 R_\Earth^2}
\left[{\mu_{\leftmoon}\over a_{\leftmoon}^3(1-e_{\leftmoon}^2)^{3/2}}\left(1-{3\over 2}\sin^2 i_{\leftmoon}\right)
+{\mu_{\odot}\over a_{\odot}^3(1-e_{\odot}^2)^{3/2}}\right].
\end{equation}
Substituting numerical values to all parameters, we have
\begin{equation}\label{eq:iminmaxnum}
i_{\min}=56.06^\circ- 0.00134\left({a\over R_\Earth}\right)^5 \, [\textrm{deg}], \,
i_{\max}=56.06^\circ+ 0.00134\left({a\over R_\Earth}\right)^5 \, [\textrm{deg}].
\end{equation}
The two red dashed curves in Fig.\,\ref{fig:fig1} mark the intersection of the two planes of fast drift corresponding to the limiting values $i_{\min},i_{\max}$ with the plane $(i,e)$ for $a =20,000$ km to provide rough delimiters of the overall extent of the resonance. More refined limits are provided by the intersections of the separatrices emanating from the unstable point $P_{0}$ under the integrable approximation, as detailed in the next subsection. Both types of curves cease to exist when chaos becomes prominent and the integrable approximation no longer holds even approximately. In practice, however, numerical evidence (see subsection \ref{sec:NHIMana}) indicates that the limiting curves of the planes of fast drift continue to represent well the borders of the resonance even in cases of strong chaos. In particular, the width of the resonance can be estimated by the difference: 
\begin{equation}\label{eq:dires}
\Delta i_\R = i_{\max}-i_{\min}
=0.00268\left({a\over R_\Earth}\right)^5\, [\textrm{deg}].
\end{equation}
Table \ref{tab:width} shows the values of $i_{\min},i_{\max}$ and $\Delta i_\R$ as found numerically by the linearization of the full Hamiltonian $\mathcal{H}_{\R}$, and by the estimates of Eqs.\,(\ref{eq:iminmaxnum}) and (\ref{eq:dires}). 

\begin{table}[h!]
	\begin{center}
		\caption{\small Minimum and maximum inclinations for which the circular orbit $P_0$ is linearly unstable, for various altitudes (values of the semi-major axis $a$). Numerical values are obtained by computing the eigenvalues of the variational matrix at $P_0$ under the full Hamiltonian $H_\R$ truncated at order $r=10$ in the actions. Analytical estimates are obtained by Eqs.\,(\ref{eq:iminmaxnum}) and (\ref{eq:dires}).} 
		\label{tab:width}
		\begin{tabular}{c|c|c|c|c|c|c}
			a (Earth radii) and in km & $i_{\min}$(deg) & $i_{\min}$(deg)& $i_{\max}$ (deg) & $i_{\max}$ (deg)& $\Delta i_\R$ (deg) & $\Delta i_\R$ (deg)\\
			~ & numerical & theoretical & numerical &  theoretical & numerical  &  theoretical \\
			\hline
			\hline
			2.97 \textrm{(19,000 km)} & 55.75 & 55.75 & 56.38 & 56.37 & 0.62 & 0.61\\
			3.76 \textrm{(24,000 km)} & 55.09 & 55.05 & 57.06 & 57.07 & 1.97 & 2.02\\
			3.99 \textrm{(25,450 km)} & 54.76 & 54.70 & 57.41 & 57.42 & 2.65 & 2.72\\ 
			4.64 \textrm{(29,600 km)} & 53.44 & 53.18 & 58.87 & 58.94 & 5.43 & 5.76\\
		\end{tabular}
		
	\end{center}
\end{table}

%-----------------------------------------
\subsection{Bifurcations and separatrices}
%-----------------------------------------

%__________________________________________________________________________
\begin{figure}
	\centering
	\includegraphics[width=1\textwidth]{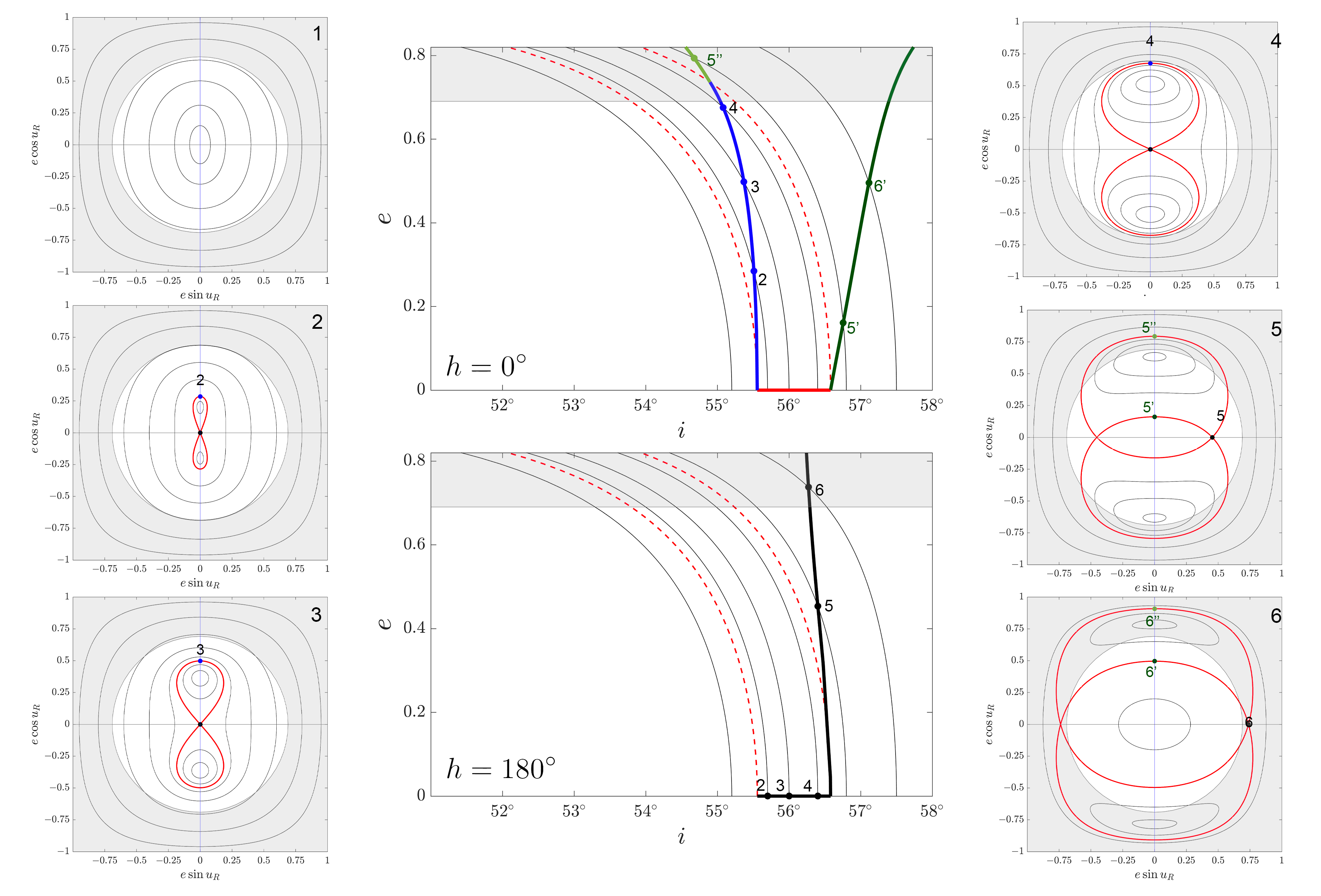}
	\caption{Sequences of bifurcations and the corresponding phase portraits in the integrable model (\ref{eq:H1dof}), for $a = 20,000$ km, as $i_{\Circ}$ (and hence $J_{\F}$) is altered in the interval $55.5^\circ\leq i_{\Circ}\leq 57.5^\circ$. 
		The two central panels, besides showing the 
		foliation by the planes of fast drift, illustrate how the separatrices emerging through the chain of successive bifurcations of unstable fixed points intersect the plane $(i,e)$ for two different choices of the initial phases of the satellite's orbit, \ie $g=0$ and $h=0$ (middle top panel), or $h=180^\circ$ (middle bottom panel).
		Six different phase portraits are shown corresponding to six values of $i_{\Circ}$. Numbered points show the correspondence between a point on one of the separatrices (in the phase portrait) and the corresponding point of intersection of the separatrix with the $(i,e)$ plane. Grey regions indicate regions below Earth's surface.}
	\label{fig:fig1}
\end{figure}
%----
The number of real and physically admissible solutions (solutions with an eccentricity smaller than the critical value $e_{\textrm{re-entry}}$) of fixed points arising in phase space around the point $P_0$ are obtained by the real roots of the equations
\begin{align}\label{eq:FP}
\left\{
\begin{aligned}
&\dot{X}= \partial_{Y}\mathcal{H}_{\R}= 0,\\ 
&\dot{Y}=-\partial_{X}\mathcal{H}_{\R}=0.
\end{aligned}
\right.
\end{align}
These depend on the specific value of the label $J_{\F}$. In an open neighborhood around the central fixed point $P_0=(0,0)$, the sequence of bifurcations of new fixed points is determined by the truncation $H_{2,4}$ of the Hamiltonian (\ref{eq:H1dof}) quadratic and quartic in the variables $X,Y$. Keeping the leading parts of all relevant coefficients yields 
\begin{equation}\label{eq:H4}
H_{2,4} = c_{02}(Y^2-X^2) 
+ c_{102}\sqrt{\mu_E a}(\cos i_\star-\cos i_{\Circ})(Y^2+X^2) 
-c_{04}(X^2+Y^2)^2,
\end{equation}
with 
\begin{equation}\label{eq:c04}
c_{04}={3J_2R_\Earth^2(5-8\cos i_{\star})\over 64a^4}.
\end{equation}
We have $c_{02}>0$, $c_{102}>0$, $c_{04}>0$, and $c_{02}=\mathcal{O}\left(a^{3/2}\mu_b/(\sqrt{\mu_\Earth}a_b^3)\right)$ (the index `b' standing for Moon or Sun), and $c_{102}=\mathcal{O}\left(J_2\mu_\Earth^{1/2}R_\Earth^2/a^{7/2}\right)$, $c_{04}=\mathcal{O}\left(J_2 R_\Earth^2/a^4\right)$. 

\begin{rmk}
	The Hamiltonian $H_{2,4}$ is sufficient for i) a qualitative discussion of the types of phase flow appearing as a result of the sequence of bifurcations taking place as $J_\F$ is altered, and ii) estimates to the quantities $e_{\max}$ and the folding time $T_e$ related to the mechanism of `eccentricity growth'. Terms of higher degree determine the exact position of the fixed points. In practice, the entire phase portrait obtained by the level curves of (\ref{eq:H1dof}) is stabilized at a truncation $N=6$ in the action variables $(J_{\F},J_{\R})$. The actual phase portraits shown in Fig.\,\ref{fig:fig1} were obtained at the truncation $N=10$.
\end{rmk}

%=======
\begin{figure}
	\centering
	\includegraphics[width=1\textwidth]{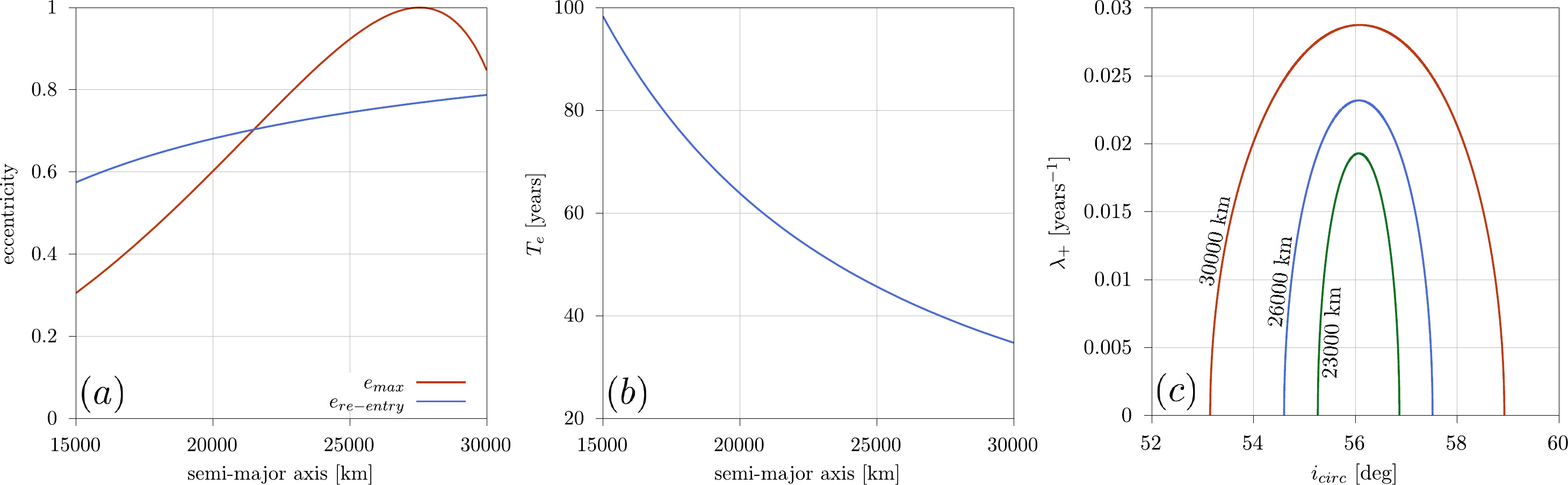}
	\caption{(a) 
		The maximum eccentricity attained for orbits with initial conditions along the separatrix of the central fixed point $P_0$ as a function of the semi-major axis $a$ (red curve, analytical expression given by Eq.\,(\ref{eq:emax})). The blue curve shows the eccentricity for which the orbit's perigee is equal to the Earth's radius as a function of $a$. (b) The e-folding time $T_e$ given by Eq.\,(\ref{eq:te}), determining the exponential rate of eccentricity growth as a function of $a$. (c) Dependence of the positive eigenvalue $\lambda_+$ of $P_0$ on $i_\Circ$ for three different altitudes defined by $a=23,000$ km (black), $a=26,000$ km (blue) and $a=30,000$ km (green). }
	\label{fig:fig2}
\end{figure}
%----------------
Fig.\,\ref{fig:fig1} presents the changes in the form of the phase portrait of the model (\ref{eq:H1dof}) at $a=20,000$ km, as the value of $i_{\Circ}$ changes between $i_{\Circ}=55^{\circ}$ and $i_{\Circ}=57.5^{\circ}$. 
Given the definition of $(X,Y)$ in Eq.\,(\ref{Eq:ResCoord}), one has $(X,Y)=(\mu_\Earth a)^{1/4}(e\sin u_\R,e\cos u_R) + \mathcal{O}(e^2)$. Thus, to facilitate insight, the phase portraits are presented in the Keplerian element variables $(e\sin u_\R, e\cos u_\R)$, $u_{\R}=-(g+h/2)$.

The phase flow type of Fig.\,\ref{fig:fig1} are in agreement with the schematic figure presented by \cite{sBr01-2} for the \textit{Second Fundamental Model for second-oder resonances} (SFM2). The  portraits highlight the changes as the circular orbit turns from stable to unstable. The transition takes place at $i_{\Circ}=i_{\min}$ given by Eq.\,(\ref{eq:iminmax}).  After that point, the figure-eight separatrix associated with the unstable origin grows in size, reaching its maximum size at $i_{\Circ}=i_\star$. From the fourth order truncation of (\ref{eq:H1dof}) in $X,Y$, the maximum eccentricity reached along the separatrix is attained for $X=0$, $Y^2 = c_{02}/c_{04}$, leading to:
\begin{equation}\label{eq:emax}
e_{\max}= \left\{1-\left[1-\left(
{5a^5(1+\cos i_\star)\sin 2\varepsilon\over (5-8\cos i_{\star})\mu_\Earth J_2 R_\Earth^2}
\left({\mu_{\leftmoon}\left(1-{3\over 2}\sin^2i_{\leftmoon}\right)\over a_{\leftmoon}^3(1-e_{\leftmoon}^2)^{3/2}}
+{\mu_{\odot}\over a_{\odot}^3(1-e_{\odot}^2)^{3/2}}\right) \right) \right]^2\right\}^{1/2},
\end{equation}
or, after valuation of the parameters, 
\begin{equation}\label{eq:emaxnum}
e_{\max}=\sqrt{1-\left(1-(6.65\times 10^{-4})\left({a\over R_\Earth}\right)^5 \right)^2}. 
\end{equation}
The left-panel of Fig.\,\ref{fig:fig2} shows the maximum eccentricity attained by orbits with initial conditions close to the figure-eight separatrix as a function of the semi-major axis for $15,000\textrm{~km}\leq a\leq 30,000\textrm{~km}$. Superposed is the curve $e_{\textrm{re-entry}}= 1-R_{\Earth}/a$, which shows the limiting eccentricity for re-entry in the Earth's atmosphere at perigee, as a function of $a$. The two curves intersect at $a\approx 23,000$ km, implying that motions close to the theoretical separatrices of the resonance can lead to re-entry\footnote{As discussed by \cite{iGk19}, the chaotic nature and the sensitivity on the initial conditions of the orbit solutions leading to eccentricity growth is manifold-drive and determined by the stable and unstable manifolds of the normally hyperbolic invariant manifold $\mathcal{C}_{\NHIM}$, as  discussed in the sequel. Still, it is numerically verified that  Eq.\,(\ref{eq:emaxnum}) gives the correct estimate for the efficiency of the mechanism, \ie the maximum eccentricity expected for the chaotic orbits with initial conditions close to the theoretical separatrices of the integrable model (\ref{eq:H1dof}).}. The eccentricity growth along the unstable direction of $P_0$ grows exponentially in time, with an e-folding time $T_e$ possible to estimate by the inverse of the positive real eigenvalue of $P_0$ at $i_{\Circ}=i_\star$. The eigenvalue $\lambda_+$ scales as
\begin{equation}\label{eq:lam}
\lambda_+ \approx 2\left(c_{02}^{2}-c_{102}^{2}J_\F^{2}\right)^{1/2} = 
2\big(c_{02}^{2}-c_{102}^{2}\sqrt{\mu_\Earth a}(\cos i_\star-\cos i_{\Circ})\big)^{1/2}.
\end{equation}
The right-panel of Fig.\,\ref{fig:fig2} shows the variation of $\lambda_+$ with $i_{\Circ}$ for three different values of the semi-major axis. The maximum, obtained for $i_{\Circ}=i_\star$, yields an e-folding time
\begin{equation}\label{eq:te}
T_e = \left({1\over \lambda_+}\right)_{i_{\Circ}=i_\star}={1\over 2}\left[{15a^{3/2}(1+\cos i_\star)\sin 2\varepsilon\over 32\mu_E^{1/2}}\left(
{\mu_{\leftmoon}(1-{3\over 2}\sin i_{\leftmoon}^2)\over a_{\leftmoon}^3(1-e_{\leftmoon}^2)^{3/2}} +
{\mu_{\odot}\over a_{\odot}^3(1-e_{\odot}^2)^{3/2}}\right)\right]^{-1/2},
\end{equation}
or 
\begin{equation}\label{eq:te2}
T_e \approx 354.43 \times \left({R_\Earth\over a}\right)^{3/2} [\textrm{yr}].
\end{equation}
One finds $T_e\approx 35$ yr at Galileo altitude (see middle-panel of Fig.\,\ref{fig:fig2}). \\

Returning to the sequence of bifurcations of Fig.\,\ref{fig:fig1}, two new unstable fixed points are generated along the $X$-axis as $i_\Circ$ grows beyond the second critical value $i_{\max}$. The transition in the associated phase portraits is shown in panels $4$ and $5$. Note that, as $i_\Circ$ increases, moving away from the resonance, the unstable points on the $X$-axis move away from the origin (panel $6$), until eventually reaching the re-entry circle at $(\mu_\Earth a)^{1/4}(1-R_\Earth/a)$. Then, the phase space becomes  foliated once again by rotational invariant curves only. \\

The two middle panels of Fig.\,\ref{fig:fig1} show, now, how to use the above information in order to compute analytically the form of the separatrices of the $\mathcal{R}_{2g+h}$ resonance on the plane $(i,e)$ for fixed value of the semi-major axis. We consider two theoretical sets of initial conditions, in which we start with $g=0$ while the initial phase $h$ is selected as $h=0$ (top panel) or $h=180^\circ$ (bottom panel). The corresponding values of the resonant angle are $u_R=0$ and $u_R=\pi/2$, implying that the choice of initial conditions is along the $Y-$axis  or $X-$axis respectively. The points $2$, $3$, $4$, $5''$ and $6''$ correspond to the uppermost point of the separatrix generated at the bifurcation at $i_\Circ=i_{\min}$, and they all have $X=0$, implying a phase $h=0$. Thus, at all these points we have $e=e_{\max}$ (confer Eq.\,(\ref{eq:emax})), while the points belong to different planes of fast drift labelled by different values of $i_\Circ$, given by the point of intersection of each plane of fast drift with the axis $e=0$. Then, the inclination can be computed by Eq.\,(\ref{eq:jf2}), setting $e=e_{\max}$, $J_\F = \sqrt{\mu_E a}(\cos i_\star-\cos i)$.    
The curve joining the points $2, 3, 4, 5'', 6''$ thus represents the intersection of the whole family of separatrices (labeled by $i_\Circ$) with the plane $(i,e)$, for the choice of phases $g=0$, $h=0$. Working in the same way, we can obtain the curve joining the points 5',6', which represents the intersection of the same plane with the family of separatrices associated with the second bifurcation (for $i_\Circ>i_{\max}$). On the other hand, the first family of separatrices produces no points for $X\neq 0$, while the second family terminates at the unstable fixed points $5, 6$. Thus, the curve joining the points $5$ and $6$ in the plane $(i,e)$ for $g=0$, $h=180^\circ$ represents, in the same plane, the family of unstable points (for different planes of fast drift) generated for $i_{\Circ}>i_{\max}$. 

The curves yielding the intersection of the separatrices with the plane $(i,e)$ provide a more precise estimate of the borders of the resonance than the estimate based on the planes of fast drift passing through the interval $\Delta i_{\R}$. However, such a definition holds only when chaos is limited, so that the chaotic layers generated at the resonance are thin and remain close to the theoretical separatrices. 

%In the next section we return to the problem of how to properly define the borders of the resonance in comparison with the results of numerical (FLI) stability maps obtained in the framework of the complete secular Hamiltonian model. 

% Repetition:

%\begin{remark}[On the reduction to the \textit{First Fundamental Model} (FFM) performed in \cite{jDa16}]	In \cite{jDa16}, all the isolated lunisolar secular resonances have been reduced to the   \textit{First Fundamental Model} model (a nonlinear pendulum) to compute various resonant widths to advocate Chirikov's superposition of resonance as the main driver of chaos for the MEO region. 
%As we just described, the reduction of the $\mathcal{R}_{2g+h}$ to the FFM is inaccurate. Its phase flow type is given by the SFM2 model, which also allows to obtain the correct form of the separatrices and the borders of the resonance in the plane $(i,e)$.  
%\end{remark}

%======================================================
\section{Dynamics of the full problem}\label{sec:NHIM}
%======================================================
The parametric integrable 1-\DOF approximation built in section \ref{sec:IntApproximation}  has been obtained under the $\mathcal{R}_{2g+h}$ isolated resonance hypothesis and treating $J_{\F}$ as a constant.  Thus, this picture is unable to explain two important features observed  in past numerical works \citep{eAl16,iGk16,aRo17}: 
\begin{enumerate}
	\item the dependence of the resonant structures depicted in stability maps on the initial phases $h$ and $\Omega_{\leftmoon}$,
	\item the onset of a large degree of chaos at the resonance crossings of the $2g+h$ resonance with other secular resonances.
\end{enumerate}
In the present section we address these phenomena on the basis of an invariant manifold structure which exists in the vicinity of the $2g+h$ resonance. 
We show how to properly define the borders of the resonance in comparison with  numerical stability maps obtained with the complete secular Hamiltonian model. 
In particular, we show how this structure is scuplted by the invariant manifold of circular orbits, which becomes a \textit{normally hyperbolic invariant manifold} (NHIM) in the vicinity of the $2g+h$ resonance (see \cite{sWi13} and references therein for the definitions related to the NHIM). Most phenomena related to hyperbolicity can be interpreted by the dynamics of the stable and unstable asymptotic manifolds of the NHIM. 

%----------------------
\subsection{The centre manifold}\label{Dyn:CM}
%----------------------
Since $\dot{X}=\dot{Y}=0$ for circular orbits whatever the values of $(i,h,\Omega_{\leftmoon})$, one obtain the  $4$-dimensional set $\mathcal{C}$,
\begin{align}
\mathcal{C}=\Big\{
(X,Y,J_{\F},u_{\F},J_{{\leftmoon}},u_{{\leftmoon}})
\,
\vert 
\,
X=Y=0 
\Big\},
\end{align}
is invariant for the flow $\Phi^{t}_{\mathcal{H}}$ under the averaged Hamiltonian of Eq.\,(\ref{eq:final_ham}).
We refer to this set as the \textit{centre manifold} of the phase space of the Hamiltonian flow. The dynamics restricted on the centre manifold is given by the restriction of the full Hamiltonian 
on $\mathcal{C}$, given by setting  $X=Y=0$:
\begin{align}\label{eq:HCM}
\mathcal{H}_{\CM}
\equiv
\eval{\mathcal{H}}_{\mathcal{C}}
%\mathcal{H}_{\big|\mathcal{C}}
=
\mathcal{H}(0,0,J_{\F},J_{{\leftmoon}},u_{\F},u_{{\leftmoon}}).
\end{align} 

%-----------------
\subsection{Splitting of the Hamiltonian}\label{Dyn:split}
%-----------------
The existence of the invariant set $\mathcal{C}$ leads to a new splitting of Eq.\,(\ref{eq:final_ham}) as
\begin{align}\label{eq:HNewSplitting}
\mathcal{H}=\mathcal{H}_{\R,0}(X,Y) + \mathcal{H}_{\CM}(J_{\F},J_{{\leftmoon}},u_{\F},u_{{\leftmoon}}) + \mathcal{H}_{\C}(X,Y,J_{\F},J_{{\leftmoon}},u_{\F},u_{{\leftmoon}}).
\end{align}
The function $\mathcal{H}_{\R,0}(X,Y)$ contains terms depending only on the resonant Poincar\'{e} variables $(X,Y)$ and reads to the lowest order
\begin{equation}\label{eq:HR0}
\mathcal{H}_{\R,0}=
\sum_{i=1}^{3} c_{(2i)0} X^{2i} + \sum_{i=1}^{3} c_{0(2i)} Y^{2i}
+ c_{22}X^2Y^2 + c_{42}X^4Y^2 + c_{24}X^2Y^4+\ldots~~.
\end{equation}
The function $\mathcal{H}_{\C}$, on the other hand, contains all terms coupling one or more of the center-manifold variables $(u_\F,u_{\leftmoon},J_\F,J_{\leftmoon})$ with the resonant variables $(X,Y)$. A crucial remark is that $\mathcal{H}_{\C}$ contains the terms 
\begin{align}\label{eq:HCoupling}
c_{120}J_\F X^2 + c_{102} J_\F Y^2 + c_{122}J_{\F}X^2Y^2 +c_{220}J_{\F}^{2}X^{2} +  
c_{202}J_{\F}^{2}Y^{2} +c_{140}J_{\F}X^{4}+c_{104}J_{\F}Y^{4}+\ldots  \notag 
\end{align}
of the integrable model (\ref{eq:H1dof}). In the previous section, we have explained how considering $J_{\F}$ as a constant parameter allows to parameterize the sequence of bifurcations influencing the stability of the fixed point $(X,Y)=0$ as well as the structure of the separatrices around it. We now develop a theory, based on an appropriate averaging of the coupling terms $\mathcal{H}_{\C}$ with respect to the angles $(u_\F,u_{\leftmoon})$, which gives rise, in the complete dynamics, to terms of the form 
\begin{align}
\tilde{c}_{120}\JFAVE X^2 + \tilde{c}_{102} \JFAVE Y^2 + \tilde{c}_{122}\JFAVE X^2Y^2 +\tilde{c}_{220}\JFAVE^{2}X^{2} +  \\ \notag 
\tilde{c}_{202}\JFAVE^{2}Y^{2}  +
\tilde{c}_{140}\JFAVE X^{4}+\tilde{c}_{104}\JFAVE   Y^{4}+\ldots \notag 
\end{align}
\ie analogous to those of the integrable resonant model (\ref{eq:H1dof}), but taking into account the \textit{time variations of $J_{\F}$} induced by the dynamics on the center manifold. The parameter $\JFAVE$ represents an effective constant of motion which can be used for the correct calculation of the change of stability of the circular orbit and of the separatrices of the resonance, as a function of the inclination, in the complete model. In the nomenclature of Celestial Mechanics, the parameter $\JFAVE$ is associated to the satellite's \textit{proper inclination}, as explained in detail below. 

%===================
\subsection{Dynamics on the center manifold far from and close to the crossings with the resonances 
	$\mathcal{R}_{h-\Omega_{{\leftmoon}}}$ and $\mathcal{R}_{2h-\Omega_{{\leftmoon}}}$}\label{sub:CMdyn}
\begin{figure}
	\center    
	\includegraphics[width=1\textwidth]{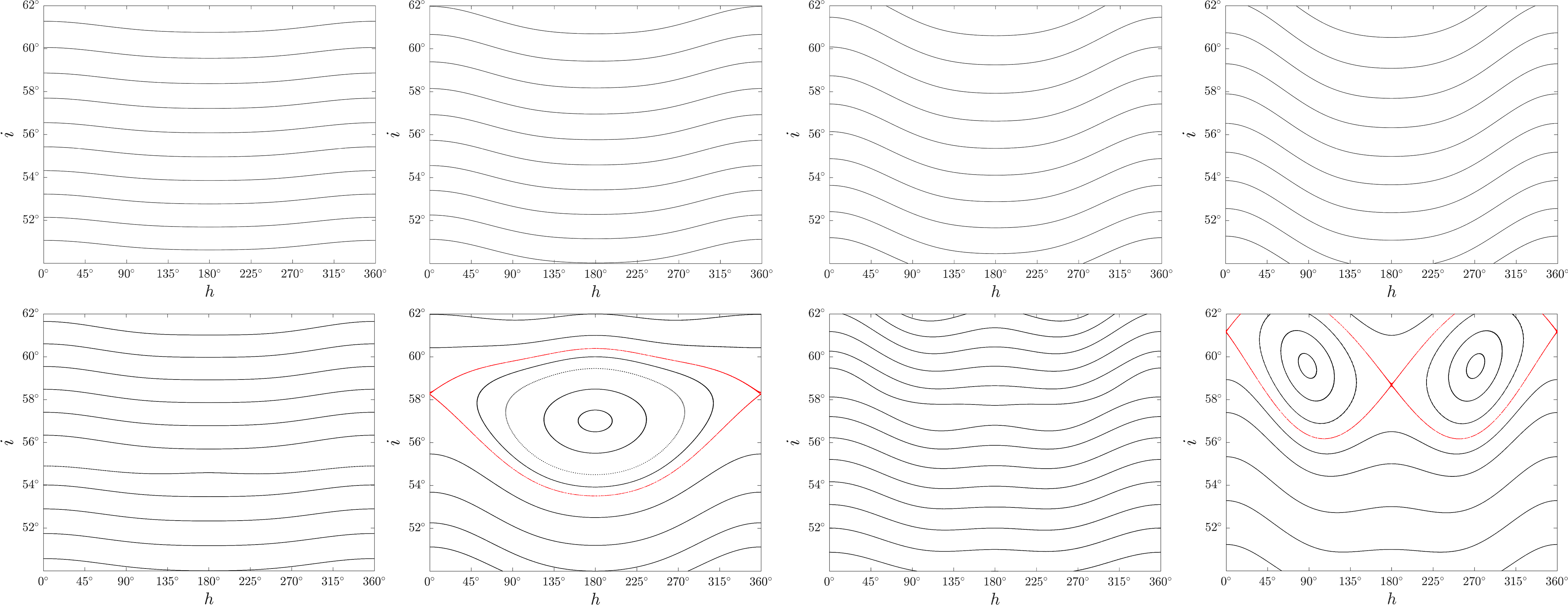}
	\caption{\label{fig:centerpc} 
		The top line shows the integrable dynamics of $\mathcal{H}_{\CM}^{(0)}$ for the values $a \in \{20,24,27,29\} \times 10^{3}$. The bottom line, for the same values of $a$, depicts the stroboscopic map, iterated $5,000$ times, associated to $\mathcal{H}_{\CM}$ and highlights the effects of 
		resonances involving the lunar node. 
	}
\end{figure}

%=================
Consider the splitting
\begin{align}\label{eq:SplitHCM}
\mathcal{H}_{\CM}(J_{\F},J_{{\leftmoon}},u_{\F},u_{{\leftmoon}})=
\mathcal{H}_{\CM}^{(0)}(J_{\F},u_{\F}) +
\mathcal{H}_{\CM}^{(1)}(J_{\F},J_{{\leftmoon}},u_{\F},u_{{\leftmoon}}),
\end{align}
where
\begin{align}
\left\{
\begin{aligned} 
& \mathcal{H}_{\CM}^{(0)}=
b_{10}J_{\F} + b_{20}J_{\F}^2 +
\sum_{i,j} b_{ij}(J_{\F})^{i}\cos(j u_{\F}), \\
& \mathcal{H}_{\CM}^{(1)}=\varpi_{\Omega_{\leftmoon}}J_{{\leftmoon}} +
\sum_{i,k_{1},k_{2}} b_{ik_{1}k_{2}} (J_{\F})^{i} \cos(k_{1}u_{\F}+k_{2}u_{{\leftmoon}}).
\end{aligned}
\right.
\end{align}
An expression for $\JFAVE$ can be arrived at by  analysing the dynamics on $\mathcal{C}$. Since the dynamics on $\mathcal{C}$ depends periodically on $u_{{\leftmoon}}$, 
its 2D stroboscopic mapping is obtained by recording 
the states $(J_{\F},u_{\F})$ at every $T_{\Omega_{\leftmoon}}$.  
Recalling that on $\mathcal{C}$ we have $e=0$ (implying $P=0$), we find 
\begin{equation}\label{eq:omeiujf}
i = \arccos\left({J_{\F}\over\sqrt{\mu_\Earth a}}+\cos i_{\star},\right), \, h=-u_{\F}, 
\end{equation}
the stroboscopic maps are directly depicted in the plane $(h,i)$ representing the values of the satellite's orbital elements at times where the stroboscopic condition $\Omega_{\leftmoon}=0$ is satisfied. 
Top line of Fig.\,\ref{fig:centerpc} shows the phase space of 
$\mathcal{H}_{\CM}^{(0)}$, the bottom line panel shows
$5,000$ iterations of the stroboscopic maps for values 
$a \in \{20,24,27,29\} \times 10^{3}$ km associated 
to  $\mathcal{H}_{\CM}$. 
In all cases, we note the large measure of KAM tori
and near complete absence of chaos. 
The tori are mostly rotational at $a=20,000$ km and $a=27,000$ km, while for $a=24,000$ km or $a=29,000$ km there exist two pendulum-like domains filled with librational tori stemming from $\mathcal{H}_{\CM}^{(1)}$ and secondary resonances involving the lunar node. These features can be understood by the following analysis.  

%----------------------
%----------------------------

%------------
\subsubsection{Inclination dynamics around the Laplace plane} 
%-----------
Far from resonances involving the argument of the lunar node $\Omega_{\leftmoon}$, by one perturbative step we can eliminate the angle $\Omega_{\leftmoon}$ from Eq.\,(\ref{eq:SplitHCM}), leading to a dynamics on $\mathcal{C}$ dominated by the $\mathcal{H}_{\CM}^{(0)}$. The dynamics is well encapsulated in the lowermost order terms of $\mathcal{H}_{\CM}^{(0)}$ whose expression is
\begin{align}\label{eq:HCM0anal}
\mathcal{\tilde{H}}_{\CM}^{(0)}=
b_{10}J_{\F} + b_{20}J_{\F}^{2}
+b_{01}\cos(u_{\F}) + b_{02}\cos(2u_{\F}).
\end{align} 
Ignoring a small variation of the frequency due to the quadratic terms $b_{20}J_{\F}^{2}$, the angle $u_{\F}$ evolves quasi-linearly with time as $u_{\F}(t) \approx u_{\F}^{0}+\omega_{\F} t$, with $\omega_{\F}$ equal to $-\dot{h}$ 
for circular orbits, given by
\begin{equation}\label{eq:omef}
\omega_{\F}\simeq b_{10}= 
\frac{3 J_2 n R_{\oplus}^2 c_{i_\star}}{2 a^2} + \frac{ 3 \mu_{\odot} c_{i_\star} (2 - 3 s_\epsilon^2)}{
	8 n r_{\odot}^3} + \frac{3 \mu_{{\leftmoon}} c_{i_\star} ( 3  s_{i_{{\leftmoon}}}^2 - 2) ( 3 s_\epsilon^2 - 2)}{16 n r_{{\leftmoon}}^3},
\end{equation}
or
\begin{equation}\label{eq:omefnum}
\omega_{\F}\simeq \left( 35.468 \left({R_\Earth\over a}\right)^{7/2} 
+ 0.001 \left({a\over R_\Earth}\right)^{3/2} \right) \,
[\textrm{yr}^{-1}].
\end{equation}
Then, from Hamilton's equation $\dot{J}_{\F}\approx-\frac{\partial \mathcal{\tilde{H}}_{\CM}^{(0)}}
{\partial u_{\F}} =  b_{01} \sin u_{\F}+2 b_{02} \sin 2 u_{\F}$
we arrive at
\begin{align}
J_{\F}(t) \approx J_{\F}^0+\int_{0}^{t} b_{01}\sin u_{\F}(s) \, \dd s
+
\int_{0}^{t} 2 b_{02} \sin 2 u_{\F}(s) \, \dd s,
\end{align}
yielding
\begin{align}\label{eq:jft}
\left\{
\begin{aligned}
&J_{\F}(t)\approx 
J_{\F}^0+\frac{b_{01}}{\omega_{\F}} 
\big(\cos u_{\F}^{0}-\cos u_{\F}(t)\big)
+\frac{b_{02}}{\omega_{\F}} 
\big(\cos 2 u_{\F}^{0}-\cos 2 u_{\F}(t) \big), \\
&u_{\F}(t) \approx u_{\F}^{0}+\omega_{\F} t.
\end{aligned}
\right.
\end{align}

Eq.\,(\ref{eq:jft}) describes oscillations in the satellite's inclination, caused by the forced inclination of the Laplace plane with respect to the Earth's equator as a result of the Moon's and Sun's inclined geocentric orbits. Substituting Eq.\,(\ref{eq:jft}) into Eq.\,(\ref{eq:omeiujf}) and expanding up to first order in the small quantity $J_{\F}$, we arrive at an expression of the inclination of a circular satellite as a function of the longitude of the ascending node valid in the whole neighborhood of $\mathcal{R}_{2g+h}$: 
\begin{equation}\label{eq:diome}
i(h) = i_p
+{c_\epsilon s_\epsilon a^5\over 2\mu_\Earth J_2 R_\Earth^2}
\left({\mu_\odot\over r_\odot^3}+{\mu_{\leftmoon}(2-3si_{\leftmoon}^2)\over 2r_{\leftmoon}^3}\right)\cos(h)
+{s_\epsilon^2 \tan(i_\star) a^5\over 8\mu_\Earth J_2 R_\Earth^2}
\left({\mu_\odot\over r_\odot^3}+{\mu_{\leftmoon}(2-3si_{\leftmoon}^2)\over 2r_{\leftmoon}^3}\right)\cos(2h).
\end{equation}
The parameter $i_p$, which depends on the initial condition $(i_0,h_0)$ as
$$
i_p = i(h_0)
-{c_\epsilon s_\epsilon a^5\over 2\mu_\Earth J_2 R_\Earth^2}
\left({\mu_\odot\over r_\odot^3}+{\mu_{\leftmoon}(2-3si_{\leftmoon}^2)\over 2r_{\leftmoon}^3}\right)\cos(h_0)
-{s_\epsilon^2 \tan(i_\star) a^5\over 8\mu_\Earth J_2 R_\Earth^2}
\left({\mu_\odot\over r_\odot^3}+{\mu_{\leftmoon}(2-3si_{\leftmoon}^2)\over 2r_{\leftmoon}^3}\right)\cos(2h_0),
$$
gives the mean value of the satellite's inclination over a complete circle of nodal precession (at the period $2\pi/\omega_F$). Thus, the parameter $i_p$ labels the invariant tori on $\mathcal{C}$, and will hereafter be referred to as the satellite's \textit{proper inclination}. Substituting numerical values for all constants, we obtain
\begin{equation}\label{eq:diomenum}
i(h) = i_p + 
\left(0.000786 \left({a\over R_E}\right)^5 \cos(h) + 0.000127 \left({a\over R_E}\right)^5 \cos(2h)\right) 
\,
[\textrm{deg}].
\end{equation}

Some theoretical curves of Eq.\,(\ref{eq:diomenum}) are superposed to the numerical invariant curves on the stroboscopic maps of Fig.\,\ref{fig:centerpc} at the semi-major axes $a=20,000$ km and $a=27,000$ km. The near exact coincidence of these curves indicates that, far from resonance crossings, the rotational tori on the center manifold just reflect the oscillations in inclination of the circular orbits induced by the shift of origin of the inclination vector at the forced equilibrium induced by the Laplace plane.     

%-------------
\subsubsection{Inclination dynamics at the crossings with the resonances 
	$\mathcal{R}_{h-\Omega_{{\leftmoon}}}$ and $\mathcal{R}_{2h-\Omega_{{\leftmoon}}}$}
\label{sub:SecRes} 
%------------
The above picture is altered drastically at points where the $\mathcal{R}_{2g+h}$ resonance crosses two other major lunisolar resonances, namely $\mathcal{R}_{h-\Omega_{{\leftmoon}}}$ and $\mathcal{R}_{2h-\Omega_{{\leftmoon}}}$, whose corresponding terms are included in $\mathcal{H}_{\CM}^{(1)}$. A careful inspection of all coefficients in the Hamiltonian (as for example $d_{11}$ and $d_{21}$; see Appendix \ref{App:FormalCoeff}) renders clear that the particular importance of these resonances stems from the fact that the corresponding coefficients scale \textit{linearly} with the Moon's relative inclination $i_{\leftmoon}$ with respect to the ecliptic as we have $d_{k1}=\mathcal{O}(\sin i_{\leftmoon})=\mathcal{O}(i_{\leftmoon})$ for $i_{\leftmoon}$ small (note that only resonances involving the argument of the lunar node fall into the previous analysis; the corresponding solar terms are absent as the Sun's orbit lies strictly on the ecliptic).   

%====================
%\begin{figure}
%\center    
%\includegraphics[width=0.7\textwidth]{figure/fig4.png}
%\caption{\label{fig:SectionHCMResonances} 
%The stroboscopic Poincar\'{e} maps of the center manifold dynamics at $a=24000$~km (left column) and $a=29600$~km (right column). The top row shows the stroboscopic phase portraits obtained under the flow of only the $\mathcal{H}_{\CM}^{(0)}$ term. The bottom row shows the phase portraits under the full center manifold Hamiltonian $\mathcal{H}_{\CM}$. The two librational domains formed in the latter case are limited by the separatrices of the $h-\Omega_{\leftmoon}$ (top) and 
%		$2h-\Omega_{\leftmoon}$ resonances (bottom). }
%\end{figure}
%===========
The resonance $\mathcal{R}_{2g+h}$  is crossed transversely by  $\mathcal{R}_{h-\Omega_{{\leftmoon}}}$ at $a=3.8\,R_{\Earth}$
and by $\mathcal{R}_{2h-\Omega_{{\leftmoon}}}$ at $a=4.69 \,R_{\Earth}$\footnote{Confer Fig.\,$5$ of \cite{iGk16} for numerical fingerprints of this crossing, slightly above Galileo's altitude.}.  
The top and bottom rows of Fig.\,\ref{fig:centerpc} show the stroboscopic Poincar\'{e} maps of the flow on $\mathcal{C}$ without and with the $\mathcal{H}_{\CM}^{(1)}$ term included in the equations of motion. We note that without the resonant terms $\cos(u_{\F}+\Omega_{\leftmoon})$ and $\cos(2u_{\F}+\Omega_{\leftmoon})$, included in $\mathcal{H}_{\CM}^{(1)}$, the dynamics looks quite similar to the one due to the oscillations in inclination induced by the presence of the Laplace plane. When the resonant terms $\cos(u_{\F}+\Omega_{\leftmoon})$ and $\cos(2u_{\F}+\Omega_{\leftmoon})$ are added, instead, the phase space is divided in two regions in which rotational or librational tori prevail. It is noticeable that, while the two resonances do not destroy the nearly completely ordered character of the orbits on $\mathcal{C}$, they have a large effect on the form of the oscillations of the inclination vector for circular orbits. For rotational tori, in particular, we find an increase of the amplitude of the oscillation by $\sim 2^\circ$ close to the separatrices of the resonances, nearly equal to the separatrix half width in both cases. The latter can be easily computed from the following simplified model
\begin{align}\label{eq:HCMGal}
\mathcal{\tilde{H}}_{\CM}^{(2,1)}=
\mathcal{\tilde{H}}_{\CM}^{(0)}
+
n_{\Omega_{\leftmoon}}J_{{\leftmoon}}
+
d_{21}\cos(2u_{\F}-u_{{\leftmoon}}),
\end{align}  
near  $\mathcal{R}_{2h-\Omega_{{\leftmoon}}}$. For the effects of the $\mathcal{R}_{h-\Omega_{{\leftmoon}}}$ resonance we use, instead, the simplified model
\begin{align}\label{eq:HCM11}
\mathcal{\tilde{H}}_{\CM}^{(1,1)}=
\mathcal{\tilde{H}}_{\CM}^{(0)}
+
n_{\Omega_{\leftmoon}}J_{{\leftmoon}}
+
d_{11}\cos(u_{\F}-u_{{\leftmoon}}).
\end{align}

%=====================
\subsection{Normal hyperbolicity of the center manifold and manifold structures}\label{sub:NHIM}
%=====================
For values of the inclination close to the critical $i_{\crit}$ value, the corresponding subset in $\mathcal{C}$ becomes normally hyperbolic, thus coming with stable and unstable manifolds of dimension $\dim(\mathcal{C})+1=5$. We describe now this manifold structure. \\

Let us first recall that an invariant submanifold $\mathcal{C}_{\NHIM} \subset \mathcal{C}$ is normally hyperbolic for the flow $\Phi^{t}$ if the following properties hold true \citep{sWi13}: 
\begin{defi}[Normally Hyperbolic Invariant Manifold]\label{def:NHIM}
	The submanifold $\mathcal{C}_{\NHIM} \subset \mathcal{C}$ is a normally hyperbolic invariant manifold for the flow $\Phi^{t}$
	if:
	\begin{enumerate}
		\item \textit{Invariance condition.} $\forall \, t, \, \Phi^{t}(\mathcal{C}_{\NHIM}) \subset \mathcal{C}_{\NHIM}$,
		\item \textit{Spliting condition.} $\forall \, x \in \mathcal{C}_{\NHIM}, \, 
		TM=T_{x}\mathcal{C}_{\NHIM} \, \oplus \, E^{s}_{x} \, \oplus \,  E^{s}_{x}$,
		\item \textit{Hyperbolicity condition.} There exist constants 
		$0 \le \beta < \mu$, $(c_{1},c_{2},c_{3}) \in \mathbb{R}^{3}$ such that
		\begin{align}
		\left\{
		\begin{aligned}
		&(x,w) \in T\mathcal{C}_{\NHIM}, \, t\in\mathbb{R}  \, \norm{D\Phi^{t}_{\mathcal{H}}(x).w} \le c_{1} e^{\beta |t|} \norm{w},  \\ \notag
		&(x,w) \in E^{s}, \, t > 0, \, \norm{D\Phi^{t}_{\mathcal{H}}(x).w} \le c_{2} e^{-\mu t} \norm{w}, \\ \notag
		&(x,w) \in E^{u}, \, t <0, \,  \norm{D\Phi^{t}_{\mathcal{H}}(x).w} \le c_{3} e^{\mu t} \norm{w}.	\notag
		\end{aligned}
		\right.
		\end{align}
	\end{enumerate}
\end{defi}	 

\begin{rmk}
	While the above definition establishes that hyperbolicity should be dominant in the directions transverse to the NHIM, the restriction of the dynamics on $\mathcal{C}_{\NHIM}$ can be quite complex with chaotic motions. Note that in Fig.\,\ref{fig:centerpc}, under our parameters, the dynamics on $\mathcal{C}$ in general (including any normally hyperbolic subset $\mathcal{C}_{\NHIM}$) is quite regular.
\end{rmk}

%-----------------------
\subsubsection{Numerical evidence of the existence of a normally hyperbolic subset of $\mathcal{C}$}
\label{sub:NHIMnum} 
%-------------------------------
For a specific value of  $u_{\leftmoon}$, we evidence the existence of normal hyperbolicity in a subset of $\mathcal{C}$ by computing finite-time Lyapunov exponents  for various initial conditions 
$(J_{\F},u_{\F})$ on the center manifold $\mathcal{C}$. Our  steps follow the procedure described by \cite{mGu09}.
For a point $x \in \mathcal{C}$, 
the tangent space $T_{x}M$ is split as
\begin{align}
T_{x}M = T_{x}\mathcal{C} \oplus T_{x}\mathcal{C}^{\textrm{orth.}},
\end{align}
where  $T_{x}\mathcal{C}^{\textrm{orth.}}$ denotes the orthogonal space of $T_{x}\mathcal{C}$.  
The tangent vectors  in each spaces are formally decomposed as
\begin{align}
\left\{
\begin{aligned}
& w \in T_{x}\mathcal{C}, \, w=(0,0,w_{J_{\F}},w_{u_{\F}},w_{J_{{\leftmoon}}},w_{u_{{\leftmoon}}}),  \\ \notag
& w \in T_{x}\mathcal{C}^{\textrm{orth.}}, \,w=(w_{X},w_{Y},0,0,0,0).
\end{aligned}
\right.
\end{align}

%--------------
\begin{figure}
	\begin{center}
		\includegraphics[width=1\textwidth]{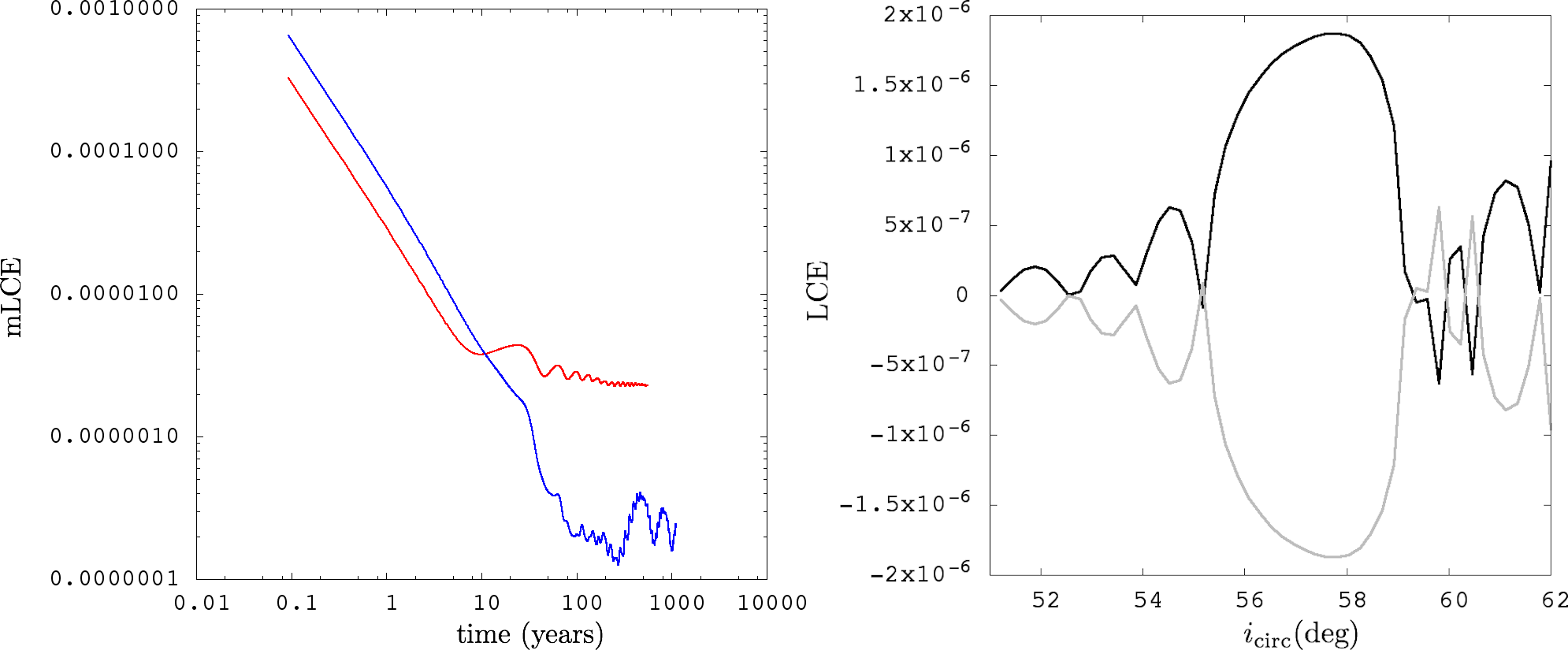}
	\end{center}
	\caption{\label{fig:TxCMortho}
		The left panel shows the prototypical mLCE associated to the $4$D space $T_{x}\mathcal{C}^{\textrm{orth.}}$ (red line) and the $2$D space
		$T_{x}\mathcal{C}$ (blue line). The stretches in the normal direction dominate the stretches tangent to the centre manifold. The right panel   reports a 
		landscape of Lyapunov exponents for $50$ initial conditions regularly spread with
		$i_{\Circ} \in [51^{\circ},62^{\circ}]$, $u_{\F}=\pi$, 
		$J_{\F}=0$ and $u_{{\leftmoon}}=0$ over $10$ Moon's nodal periods with initial tangent vectors in $T_{x}\mathcal{C}^{\textrm{orth.}}$. 
		The positive and negative exponents, coming in pair, is indicative that
		$T_{x}\mathcal{C}^{\textrm{orth.}}$ splits into a stable and unstable space.} 
\end{figure}
%------
To assess if the stretches in the normal direction dominate those tangent to the centre manifold $\mathcal{C}$, 
for initial conditions $(J_{\F},u_{\F})$, we compared the maximal Lyapunov Characteristic Exponent (mLCE) associated to the subspaces $T_{x}\mathcal{C}^{\textrm{orth.}}$ and $T_{x}\mathcal{C}$. We generically found the mLCE associated to $T_{x}\mathcal{C}^{\textrm{orth.}}$ to be larger (by more than one order of magnitude) than the mLCE associated to $T_{x}\mathcal{C}$, as shown in the left plot of Fig.\,\ref{fig:TxCMortho}.
To ensure that $T_{x}\mathcal{C}^{\textrm{orth.}}$ splits effectively into a stable and unstable direction, we computed the spectrum of the Lyapunov exponents associated to the 2D space $T_{x}\mathcal{C}^{\textrm{orth.}}$. We found Lyapunov exponents coming in pairs (with opposite sign), thus indicative of the existence of a splitting into a stable and unstable subspace, as reported in the right part of Fig.\,\ref{fig:TxCMortho}.

Another test of normal hyperbolicity refers to the computation of the characteristic multipliers for periodic orbits of the flow $\mathcal{H}_{\CM}$. These are computed  by looking at the fixed points of the associated 2D stroboscopic mapping $P_{\mathcal{H}_{\CM}}$. 
By the invariance properties of the flow on $\mathcal{C}$, the periodic orbits of the flow of $\mathcal{H}_{\CM}$ are also periodic orbits under the full Hamiltonian $\mathcal{H}$. Their stability is then computed by the eigenvalues of the associated monodromy matrix. We find numerically that all the periodic orbits associated with the elliptic fixed points in Figs.\,\ref{fig:centerpc} are \textit{simply unstable}, namely, the associated monodromy matrix has  one pair of complex conjugate eigenvalues on the unitary circle (yielding the linearized map in the tangent space of $\mathcal{C}$ at the fixed point), and one pair of real and opposite eigenvalues with corresponding eigenvectors transverse to $\mathcal{C}$. On the other hand, the saddle fixed points on the stroboscopic map correspond to doubly-unstable periodic orbits, with dominant real eigenvalues also in eigendirections transverse to $\mathcal{C}$.  

%=====================
\subsection{Limits of the $\mathcal{R}_{2g+h}$ resonance under the full Hamiltonian} \label{sec:NHIMana}
%======================
In section \ref{sec:IntApproximation} we explained how to predict the correct form of the separatrices of the $\mathcal{R}_{2g+h}$ resonance on the plane $(i,e)$ in the framework of an integrable model. We will now show how it is possible to provide meaningful analytical predictions for the width of the $2g+h$ resonance as well as to specify the limits (or separatrices) of the resonance  in the framework of the full problem.
%-----------
\subsubsection{Nearly integrable regime}\label{sec:NHIMana}
%-----------

%--------------------------
\begin{figure}
	\center    
	\includegraphics[width=0.9\textwidth]{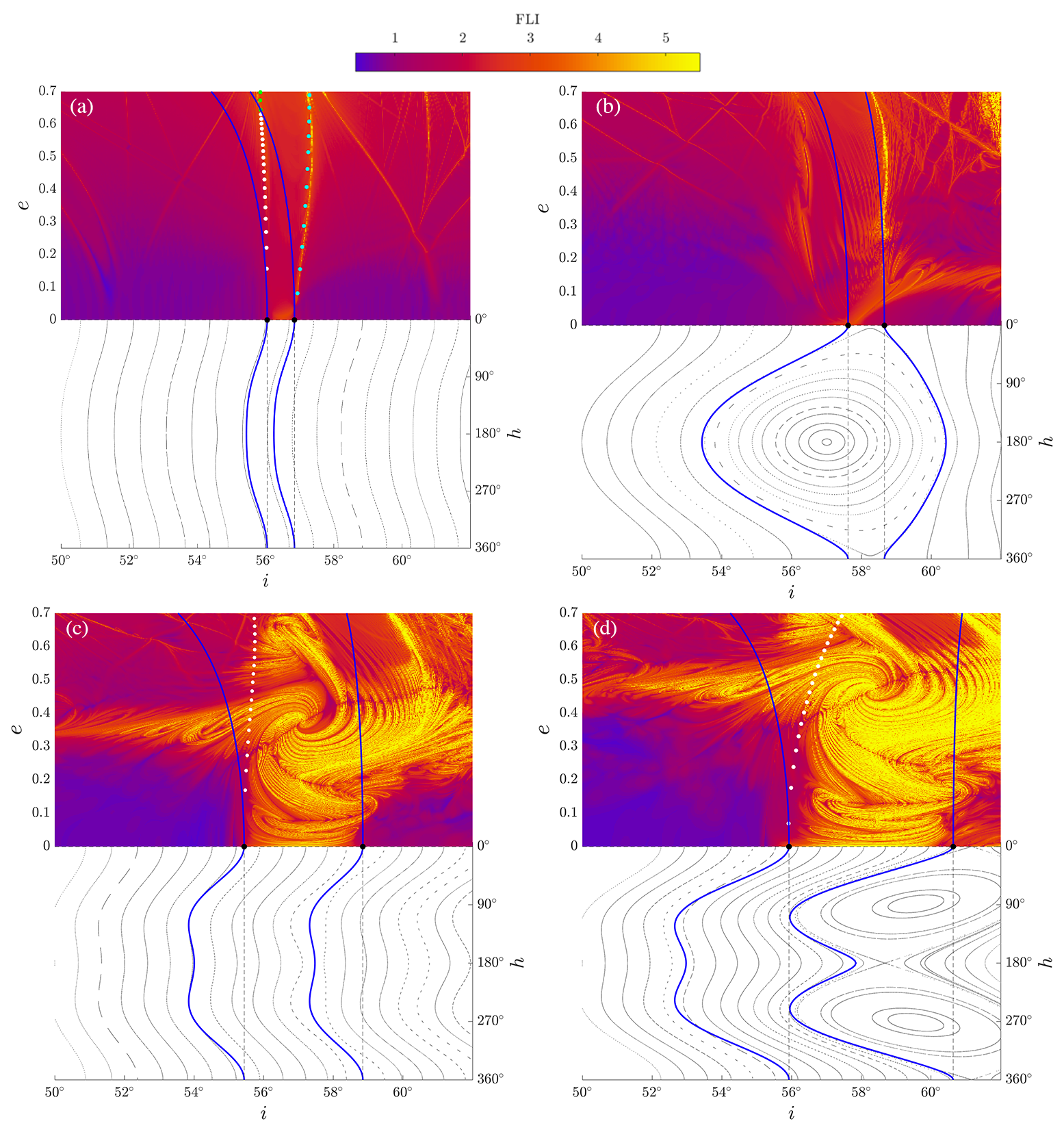}
	\caption{\label{fig:fliOme0map} 
		Comparison of the FLI stability maps with the semi-analytical predictions for the limits of $\mathcal{R}_{2g+h}$ when 
		(a) $a=20,000$ km, (b) $a=24000$ km (crossing between $\mathcal{R}_{2g+h}$ and $\mathcal{R}_{h-\Omega_{{\leftmoon}}}$), (c) $a=27,000$ km and (d) $a=29,000$ km (crossing of  $\mathcal{R}_{2g+h}$ with $\mathcal{R}_{2h-\Omega_{{\leftmoon}}}$). 
		Low values of FLIs correspond to regular orbits whilst yellow FLIs correspond to strongly chaotic orbits. 
		The initial conditions are taken in a $300 \times 300$ mesh with $g=h=\Omega_{\leftmoon}=0$ and propagated over slightly more than $16$ $T_{\Omega_{\leftmoon}}$.  
		Each FLI map comes with the stroboscopic Poincar\'{e} map associated to the center manifold Hamiltonian $\mathcal{H}_{\CM}$ (at the corresponding semi-major axis $a$). 
		The blue invariant tori indicate the limits of the subset $\mathcal{C}_{\NHIM}$, according to the adopted semi-analytical criterion of normal hyperbolicity given in Eq.\,(\ref{eq:hessian}). 
		The vertical dashed lines are tangent to the limiting tori of the NHIM at the angles $h=0$. These lines give the limiting values of the inclination $i_{\min},i_{\max}$ which substitute the estimates of Eq.\,(\ref{eq:iminmaxnum}) of the integrable 1-\DOF model of the resonance dealt with in section \ref{sec:IntApproximation}. In the FLI map, these lines project onto the limiting values of the resonance in the horizontal axis $e=0$. 
		The two blue curves in the FLI maps give the intersections of the planes of fast drift $\Pi_{\min}$ and $\Pi_{\max}$ labeled by the values of $\JFAVE_{\min}$ and $\JFAVE_{\max}$ corresponding to the inclinations $i_{\min}$ and $i_{\max}$. The thick dots in each FLI map indicate the semi-analytical separatrices of $\mathcal{R}_{2g+h}$ computed as in Fig.\,\ref{fig:fig1}, but using the model of Eq.\,(\ref{eq:HStabilityCirc}), derived from the full Hamiltonian instead of the integrable model of  Eq.\,(\ref{eq:H1dof}).}
\end{figure}
%-----------------

%-----------------
\begin{figure}
	\center    
	\includegraphics[width=0.9\textwidth]{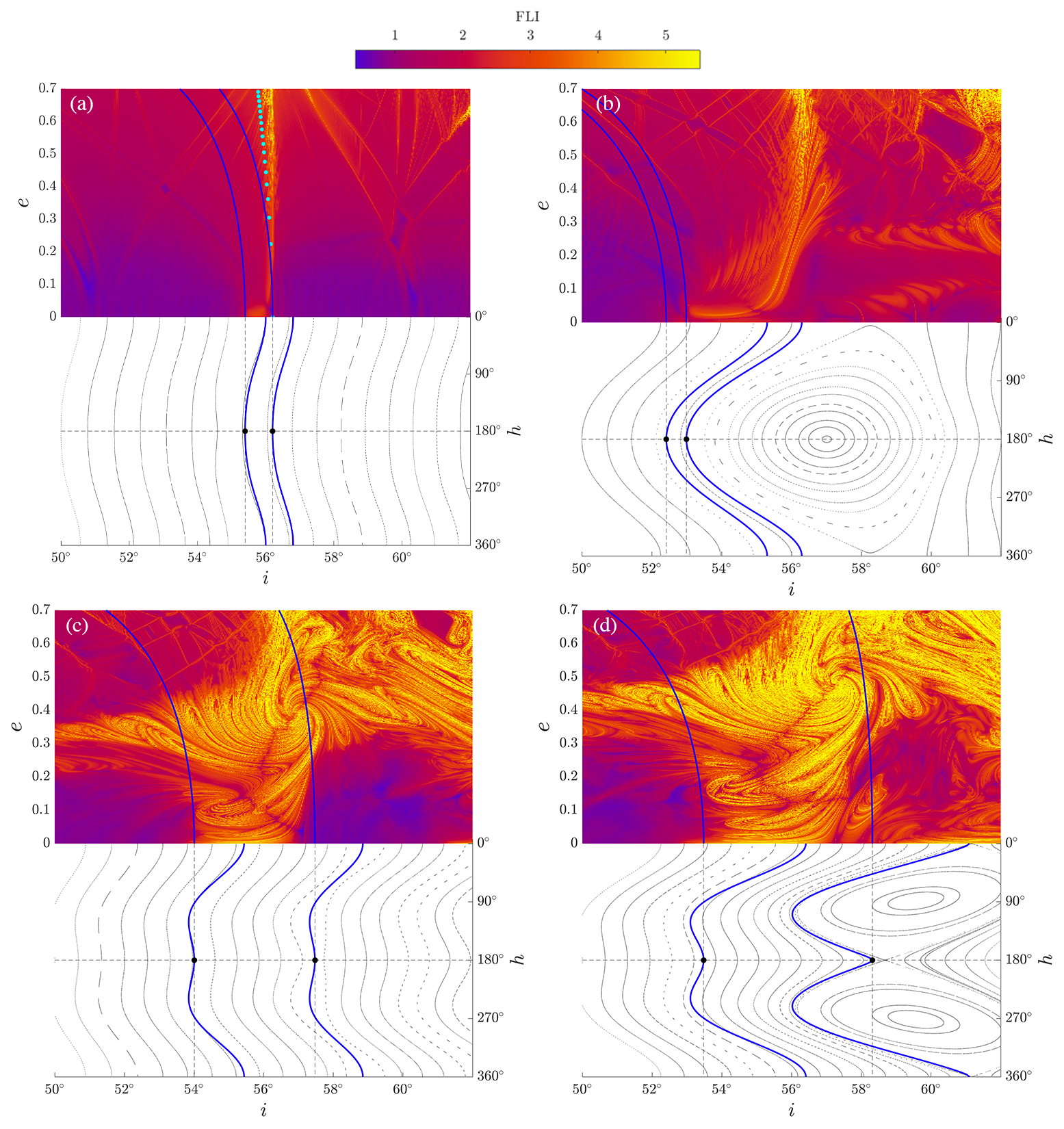}
	\caption{\label{fig:fliOme180map} 
		Same as in Fig.\,\ref{fig:fliOme0map} with $h=180^{\circ}$.}
\end{figure}
%-----------------

Fig.\,\ref{fig:fliOme0map} and \ref{fig:fliOme180map} provide the information needed for the study of the limits and structure of the resonance in the context of the full Hamiltonian model. 
The figures relate the FLI \citep{cFr00} stability maps in a portion of the $(i,e)$ plane for the 
values $a \in \{20,24,27,29\}\times10^{3}$ km with Poincar\'e maps of $\mathcal{H}_{\CM}$. 
Similar figures were produced for several values of $a$ in the range $[20,30] \times 10^{3}$ km and lead to the following observations:  
\begin{itemize}
	\item Below $23,000$ km, and similarly to the case $a=20,000$ km presented in the Figs.\,\ref{fig:fliOme0map} and \ref{fig:fliOme180map}, the general structure of the $2g+h$ resonance is as modeled in section \ref{sec:IntApproximation}. In particular, the orbits in the resonant domain are rather regular, and the resonance borders are well marked by thin separatrix-like limits appearing in yellow in the corresponding FLI maps. 
	These limits have the same qualitative structure as those computed analytically. A quantitative re-production of these curves requires, instead, a semi-analytical model stemming from the full Hamiltonian model.  
	
	\item Between $24,000$ km and $25,000$ km, $\mathcal{R}_{2g+h}$  interacts with  $\mathcal{R}_{h-\Omega_{{\leftmoon}}}$.
	This implies a complete loss in accuracy of the semi-analytical prediction on the separatrices, especially in the case $h=180^\circ$ (confer Fig.\,\ref{fig:fliOme180map}). However, estimates on the width of the resonance for small eccentricities, \ie close to the NHIM part of the center manifold $\mathcal{C}$ are still valid. 
	
	\item Between $26,000$ km and $28,000$ km,  $\mathcal{R}_{2g+h}$, $\mathcal{R}_{h-\Omega_{\leftmoon}}$ and $\mathcal{R}_{2h-\Omega_{\leftmoon}}$ interact. This leads to a strong degree of chaos inside the resonance. The limits (blue curves, see below) defined by the normal hyperbolicity condition on the subset $\mathcal{C}_{\NHIM}$ of $\mathcal{C}$ provide a rather accurate definition of the extent of  $\mathcal{R}_{2g+h}$  in the plane $i-e$.  Despite the strongly chaotic character of the motions, semi-analytical predictions based on the form of thin separatrices prove useful, to some extent, in further delimiting the resonance. 
	
	\item Above $29,000$ km, we are in the crossing domain between the $\mathcal{R}_{2g+h}$ and $\mathcal{R}_{2h-\Omega_{\leftmoon}}$. For small eccentricity values, the limits provided by the normal hyperbolicity condition on $\mathcal{C}$ still work (as for lower values of $a$). Nevertheless, due to the increasing value of the inclination of the Laplace plane, the oscillations in inclination become of several degrees, and result in a substantial overlapping between the rightmost part of the domain of $\mathcal{R}_{2g+h}$ and the leftmost part of the domain of the adjacent $\mathcal{R}_{2g}$ resonance. The overlapping between the two resonances becomes, in fact, nearly complete for $a>32,000$ km, a fact impeding further analysis by means of a semi-analytic model. 
\end{itemize}

Let us now describe how we obtain semi-analytical estimates of the borders of the resonance mentioned above. Using the parametrization $i(h;i_{p})$ for the rotational tori on the center manifold of Eq.\,(\ref{eq:diome}), we obtain 
\begin{equation}\label{eq:jfdiome}
J_{\F}(u_{\F};i_p) = \sqrt{\mu_\Earth a}\left(\cos i_\star-\cos(i(-u_{\F}))\right),
\end{equation}
which parametrizes the rotational tori of $\mathcal{C}$ in the canonical variables $(J_{\F},u_{\F})$. While strictly applicable only to circular orbits, we naively consider Eq.\,(\ref{eq:jfdiome}) as approximating the oscillations in inclination also for orbits of non-zero eccentricity\footnote{
	A more accurate model for the oscillations in inclination when $e\neq 0$ can be constructed using few steps of canonical perturbation theory, which however has a different form far from, or near the crossings with the resonances $\mathcal{R}_{h-\Omega_{\leftmoon}}$, $\mathcal{R}_{2h-\Omega_{\leftmoon}}$; see \cite{eLe21}.}. Considering the coefficients $b_{01}/\omega_F$ and $b_{02}/\omega_F$ as small quantities, Eq.\,(\ref{eq:jfdiome}) is equivalent, up to first order terms, to the equation
\begin{equation}\label{eq:jfdiome2}
J_{\F}(u_{\F};J_p) = \JFAVE-{b_{01}\over\omega_{\F}}\cos(u_{\F})-{b_{02}\over\omega_{\F}}\cos(2u_{\F}),
\end{equation}
with
\begin{equation}\label{eq:jfaveip}
\JFAVE(i_p) = 
\sqrt{\mu_\Earth a}\left(\cos i_\star-\cos(i_p)\right). 
\end{equation}
Substituting now $J_{\F}$ in the \textit{full} Hamiltonian with Eq.\,(\ref{eq:jfdiome2}), and averaging over the angles $u_{\F},\Omega_{\leftmoon}$, we arrive at
\begin{align}\label{eq:HStabilityCirc0}
\bar{\mathcal{H}}(X,Y;\JFAVE,J_{\leftmoon}) &=
\frac{1}{(2 \pi)^{2}} \int_{0}^{2 \pi} 
\int_{0}^{2 \pi} 
\mathcal{H}(X,Y,J_{\F}(u_{\F};i_p),u_{\F},J_{\leftmoon},u_{\leftmoon}) \,\dd u_{\F} \, \dd u_{{\leftmoon}},
\end{align}
yielding
\begin{eqnarray}\label{eq:HStabilityCirc}
\bar{\mathcal{H}}&=&
\tilde{b}_{10}\JFAVE + n_{\Omega_{\leftmoon}} J_{\leftmoon} +\tilde{b}_{20}\JFAVE^2 + \nonumber\\
&+&\tilde{c}_{120}\JFAVE X^2 + \tilde{c}_{102} \JFAVE Y^2 + \tilde{c}_{122}\JFAVE X^2Y^2 +\tilde{c}_{220}\JFAVE^{2}X^{2} + \tilde{c}_{202}\JFAVE^{2}Y^{2} \\
&+&\tilde{c}_{140}\JFAVE X^{4}+\tilde{c}_{104}\JFAVE Y^{4}+\cdots. \nonumber
\end{eqnarray}
The Hamiltonian $\bar{\mathcal{H}}$ is identical in structure as the Hamiltonian $\mathcal{H}_{\R}$, with the parameter $\JFAVE$, depending on the proper inclination $i_p$, in place of the integral $J_{\F}$ (which parameterizes the solutions of $\mathcal{H}_R$ for the variables 
$(X,Y)$). Thus the analysis of section \ref{sec:IntApproximation} 
can be repeated to $\bar{\mathcal{H}}$. 
%-----------------
\subsubsection{Criterion of normal hyperbolicity of the tori on $\mathcal{C}$}
%-----------------
The invariant 2D-tori of the model $\bar{\mathcal{H}}$ are labeled by the parameter $\JFAVE$, yielding the frequencies
\begin{align} 
&\dot{u_{\F}} = \tilde{b}_{10}+2\tilde{b}_{20}\JFAVE+\ldots, \notag \\
&\dot{u}_{\leftmoon}=n_{\Omega_{\leftmoon}}. 
\end{align}
These tori foliate the center manifold $\mathcal{C}$. 
The subset $\mathcal{C}_{\NHIM}\subset\mathcal{C}$ corresponds to values of $\JFAVE(i_p)$ for which 
the matrix
\begin{equation}\label{eq:hessian}
D\big(\JFAVE(i_p)\big) = \left(
\begin{array}{cc}
{\partial^2\bar{\mathcal{H}}\over\partial X\partial Y} 
&{\partial^2\bar{\mathcal{H}}\over\partial Y^2} \\
-{\partial^2\bar{\mathcal{H}}\over\partial X^2} 
&{\partial^2\bar{\mathcal{H}}\over\partial X\partial Y} 
\end{array}
\right)_{X=Y=0}
\end{equation}
has real opposite eigenvalues $\lambda_1(i_p)>0, \lambda_2(i_p)=-\lambda_1(i_p)$ is normally hyperbolic, with $\beta=0$, $\mu=\lambda_1$. 

Attached to the bottom of each FLI map in Figs. \ref{fig:fliOme0map} and \ref{fig:fliOme180map} are the numerical stroboscopic surfaces of section obtained by the center manifold Hamiltonian $\mathcal{H}_{\CM}$. The thick curves marked in blue correspond to the tori computed by Eq.\,(\ref{eq:diome}) for the limiting values of the proper inclination $i_{p,min}$, $i_{p,max}$ such that the criterion of normal hyperbolicity is met for all values of $\JFAVE(i_p)$ with $i_{p,\min}<i_p<i_{p,\max}$. Projecting the actual value of the inclination $i(h;i_p)$ on the torus labeled by $i_p$ when $h=0$ (Fig.\,\ref{fig:fliOme0map}) or $h=180^\circ$ (Fig.\,\ref{fig:fliOme180map}) we see a very good correspondence to the limits of $\mathcal{R}_{2g+h}$ for $e=0$ as specified in the corresponding FLI maps. In practice, we have that the baseline ($e=0$), as well as the whole accompanying vertical structure of the resonance is shifted to the left (lower values of the inclination) as the chosen initial phase $h$ of the trajectories increases from $0$ to $180^{\circ}$, following the corresponding change along the rotational tori of $\mathcal{C}_{\NHIM}$. This trend explains the dependence of the FLI maps on the initial phase $h$ noted by \citep{jDa16,eAl16,aRo17}. The picture is reversed symmetrically as $h$ turns from $180$ to $360^{\circ}$. 

From the points $(i=i_{\min},e=0)$ and $(i=i_{\max},e=0)$ in the $i-e$ plane emanate two curves (blue) corresponding to the intersection of the planes of fast drift $\Pi_{i_{\min}}$, $\Pi_{i_{\max}}$ with the $i-e$ plane at fixed values of the phases $h,\Omega_{\leftmoon}$. We observe that the planes of fast drift $\Pi_{i_{\min}}$, $\Pi_{i_{\max}}$ delimit reasonably well the extent of the $2g+h$ resonance as specified by the numerical FLI maps. In the case $a=20,000$ km, however, the thin separatrix-like weakly-chaotic layers marking the borders of the resonance can be approximated more precisely using the method of section \ref{sec:IntApproximation} implemented to the Hamiltonian (\ref{eq:HStabilityCirc}). This method, however, is not accurate in the case of the crossing of the $\mathcal{R}_{2g+h}$ and $\mathcal{R}_{h-\Omega_{\leftmoon}}$ resonances. Beyond that resonance, we find that the right separatrices in both Figs.\,\ref{fig:fliOme0map} and \ref{fig:fliOme180map} practically coincide with the curve $\Pi_{i_{\max}}$, while the left separatrices in Fig.\,\ref{fig:fliOme0map}, albeit produced with an integrable model, give a surprisingly precise limit of the resonance in which chaos is now strong. 

%-------------------
\subsubsection{Dependence on the initial phase $\Omega_{\leftmoon}$}
\begin{figure}
	\center    
	\includegraphics[width=0.7\textwidth]{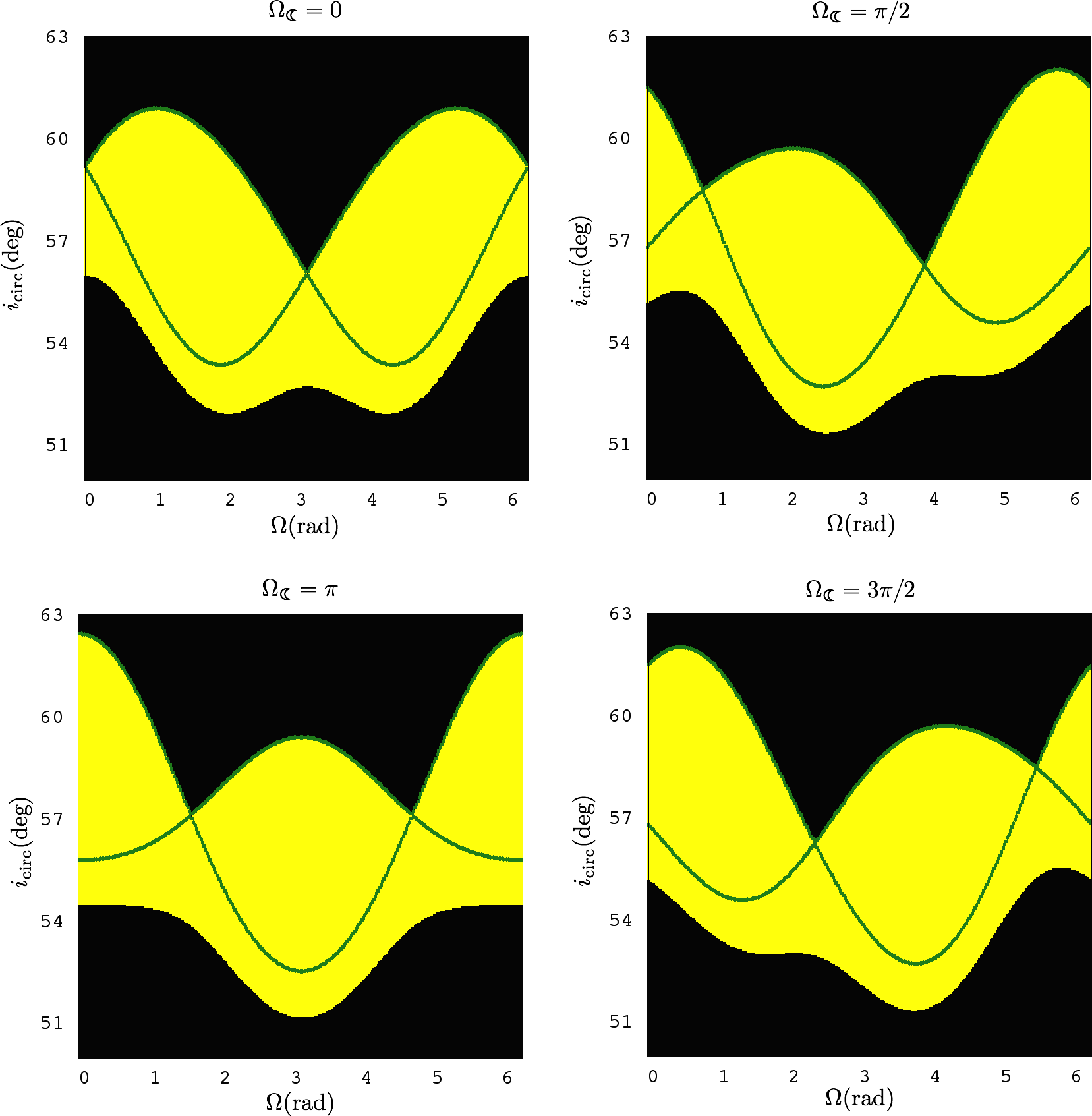}
	\caption{\label{fig:AnalyticalStability} 
		Analytical stability maps as a function of  $\Omega_{{\leftmoon}}$. The domains displayed in yellow correspond to initial points leading the Jacobian of $\bar{\mathcal{H}}$ to possess real eigenvalues for circular orbits. The green curves superimposed correspond to the separatrices of the integrable model 
		$\mathcal{H}_{\gal}$ given by Eq.\,(\ref{eq:HCMGal}). For a fixed value of $i_{\textrm{circ.}}$, both $h$ and $\Omega_{{\leftmoon}}$ influence the stability of circular orbits.}
\end{figure}

%------------------
The above semi-analytical approach introduces no dependence of the limits of normal hyperbolicity, as depicted in the FLI maps, on the choice of initial phase of the Moon $\Omega_{\leftmoon}$. Such a dependence, however, is generated close to both resonances $\mathcal{R}_{h-\Omega_{\leftmoon}}$, $\mathcal{R}_{2h-\Omega_{\leftmoon}}$ as a result of the presence of $\Omega_{\leftmoon}$ in the corresponding center manifold Hamiltonian models, confer  Eq.\,(\ref{eq:HCM11}) and  Eq.\,(\ref{eq:HCMGal}). 
The domains of normal hyperbolicity, computed with the criterion of the eigenvalues of the matrix $D\big(\JFAVE(i_p)\big)$ are shown in yellow in Fig.\,\ref{fig:AnalyticalStability}, for $a=29,600$km, and four different choices of stroboscopic section $\Omega_{\leftmoon} = 0, 90^\circ, 180^\circ$ and $270^\circ$. The domain of normal hyperbolicity in this case includes the whole domain of librational tori delimited by the separatrix of  $\mathcal{R}_{2h-\Omega_{\leftmoon}}$ (green thick line). Depending on the choice of initial phase $h$, we then see that both the position and extent in inclination of the resonance $2g+h$ in the FLI map alter with the choice of the initial phase $\Omega_{\leftmoon}$. 

%-----------------
%-----------------

%=========
\section{Stable and unstable manifolds of the NHIM. Fast escapers}\label{sec:NHIM-Escapers}
\begin{figure}
	\centering
	\includegraphics[width=1\textwidth]{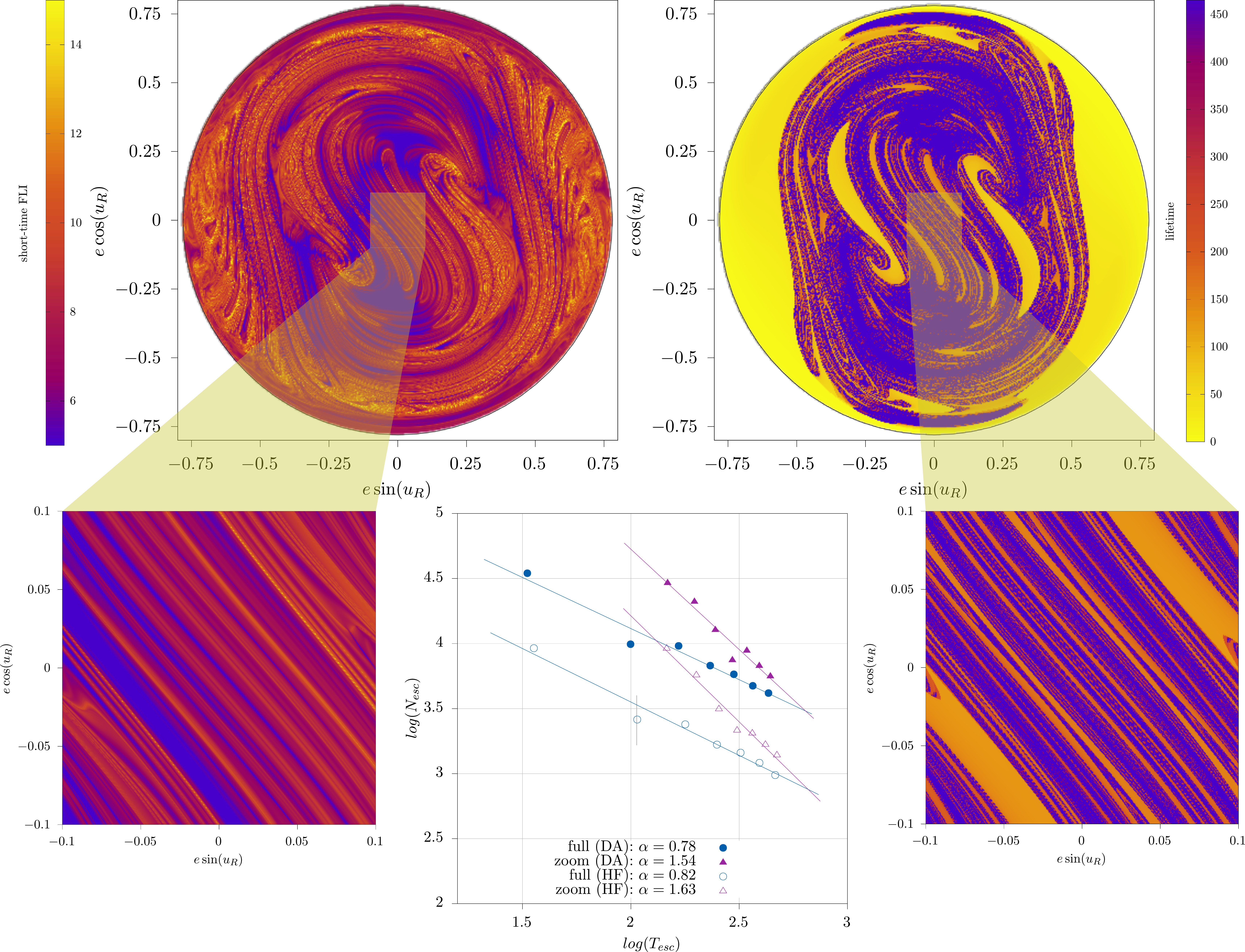}
	\caption{\label{fig:FLITESC}  
		The dynamical maps depict the stable manifold of the NHIM of the full problem $\mathcal{H}$ and the escape times $T_{\textrm{esc}}$ at two different scales of the phase space. 
		The distributions of $T_{\textrm{esc.}}$ are well approximated by power-laws. 
		The statistics derived under the secular dynamics are
		robust to the averaging principle and consistent with the non-averaged osculating dynamics.}
\end{figure}

%====
Since the subset $\mathcal{C}_{\NHIM}$ is a NHIM, the linear invariant subspaces $E^{u}$, $E^{s}$ of the tangent flow extend to stable and unstable invariant manifolds $\mathcal{W}^{s}(\mathcal{C}_{\NHIM}), \mathcal{W}^{u}(\mathcal{C}_{\NHIM})$ of $\mathcal{C}_{\NHIM}$ under the complete Hamiltonian flow. The invariant subsets $\mathcal{W}^{s}(\mathcal{C}_{\NHIM}), \mathcal{W}^{u}(\mathcal{C}_{\NHIM})$ contain trajectories which converge exponentially fast towards $\mathcal{C}$ in the forward or backward sense of time respectively. 
More precisely, the manifolds are fibrations of fibers from each point in $\mathcal{C}_{\NHIM}$
\begin{align}
\left\{
\begin{aligned}
&\mathcal{W}^{s}(\mathcal{C}_{\NHIM})=\bigcup_{x \in \mathcal{C}_{\NHIM}}\mathcal{W}^{s}(x), \\
&\mathcal{W}^{u}(\mathcal{C}_{\NHIM})=\bigcup_{x \in \mathcal{C}_{\NHIM}}\mathcal{W}^{u}(x). 
\end{aligned}
\right.
\end{align}
Those sets are efficiently revealed using FLIs computed in short-time \citep{eLe16}. The latter, computed over two scales of the phase space, are particularised to the Galileo-like dynamics ($a=29,600$ km and $i_{\textrm{circ}}=57^{\circ}$), are presented within the composite Fig.\,\ref{fig:FLITESC}. At the macroscale, we distinctively follow the lobes corresponding to the intricate structure created by the homoclinic dynamics. 
The fractal-like stratification \citep{sBl88,jNa05} appears clearly in the vicinity of $X=Y=0$. 
The practical interesting implication of the tangles generated by the stable and unstable manifold concerns the orbital lifetime of MEO objects. 
In fact, by computing the corresponding escapes times $T_{\textrm{esc}}$ maps, \ie the time taken for an orbit to reach a perigee altitude allowing an atmospheric re-entry ($\sim 120$ km for the specific choice of $a$),
we immediately note the remarkable resemblance of those maps with their FLIs counterpart (confer the right-hand side of panel \ref{fig:FLITESC}). 
We further characterised the  transport properties related to the tangle  by computing 
the distribution associated to the escapers. We found the histogram of $T_{\textrm{esc.}}$ for the region near the origin well fitted by a power-law, $P(t)\sim t^{-\alpha}$, with $\alpha \approx 1.5$, turning to $\alpha = 0.8$ at the whole phase space scale. Those laws turn to be almost independent of
of the double-averaged (DA) formulation of the dynamics. In fact, similar and compatible statistics have been derived using  a non-averaged formulation of the dynamics relying on a high-fidelity (HF) model as developed in \cite{aRo19}. Those statistics, indicative of anomalous transport \citep{rDv98,jAg01,gZa02}, are thus robust to the averaging principle and pride, one more time, the use of semi-analytical theories in satellite theory (confer central panel of Fig.\,\ref{fig:FLITESC}).  

%------------
%-----------

%======================
\section{Conclusion}
%====================== 
Based on a well accepted doubly averaged secular model, the present study revisited in-depth the dynamics of the $2g+h$ resonance. The main contributions and conclusions are the following:
\begin{enumerate}
	\item In section \ref{sec:IntApproximation}, we presented the methodology to obtain the correct form of the separatrices of $\mathcal{R}_{2g+h}$ taking into account the bifurcations associated to the parametric resonant integrable model. In particular, the method corrects the heuristic drawing of the searatrices presented by \cite{jDa16}.   
	\item In section \ref{sec:NHIM}, we have explained how meaningful estimates of the width of $\mathcal{R}_{2g+h}$ can be obtained beyond the parametric integrable resonant picture, \ie under the  whole Hamiltonian flow. The central keys in this process rely on i) the dynamical understanding associated to the invariant manifold associated to circular orbits (centre-manifold) allowing ii) to substitute the initial label of the resonant model with a new label (connected to the concept of \textit{proper inclination}). The study of the inclination dynamics has revealed how and why secondary resonances, especially including the lunar node, alter structures depicted in former numerical studies. 
	\item We provided numerical and semi-analytical evidences of the existence of a subset of the centre-manifold $\mathcal{C}$ of circular orbits turning normally hyperbolic. 
	Using the new label related to the proper inclination, analytical criteria to predict the range of normal hyperbolicity of a subset of centre-manifold have been established.
	\item Finally, the possible appearance of large scale chaos in the vicinity of the $2g+h$ resonance emanating from circular orbits has been clarified at the light of the aforementioned NHIM. The saddle structure associated with the invariant hyperbolic subset comes with stable and unstable manifolds, potentially well developed into the phase space and intersecting transversely. The chaotic dynamics steaming from those homoclinic lobes strongly induces transport and  dictates objects' lifetimes. For Galileo-like parameters, we found the distribution of those escapers to follow power-laws, whose exponents have been estimated (section \ref{sec:NHIM-Escapers}). The exponents found under the double-averaged hypothesis are in agreement with the statistics derived with a high-fidelity and osculating formulation of the dynamics. 
\end{enumerate}

%=============================
\begin{appendices}
 	
%=============================
\section{Leading terms coefficients}\label{App:FormalCoeff}
%=============================
 The formal coefficients of the leading terms appearing in the resonant Hamiltonian of Eq.\,(\ref{eq:H1dof}), the center-manifold Hamiltonian of Eq.\,(\ref{eq:HCM})  and the coupling terms of Eq.\,(\ref{eq:HCoupling}) are summarised within Tab.\,\ref{Tab:HR}, \ref{Tab:HCM} and \ref{Tab:HC} respectively. 
 The physical units adopted in this study read 
 $\mu_{\odot}=1.32712 \times 10^{11} \textrm{km}^{3}/\textrm{s}^{2}$,
 $\mu_{{\leftmoon}}=4\,902.8 \, \textrm{km}^{3}/\textrm{s}^{2}$,
 $\mu_{\Earth}=398\,600 \, \textrm{km}^{3}/\textrm{s}^{2}$,
 $J_{2}=1.082 \times 10^{-3}$, $R_{\Earth}=6\,387.1 $km, 
 $r_{\odot}=1.49579 \times 10^{8}$ km, 
 $r_{{\leftmoon}}=384\,157$ km. The shortcuts $c_{i_{{\leftmoon}}}$ and $s_{i_{{\leftmoon}}}$ 
 denote the cosine and sinus of the inclination of the Moon to the ecliptic $i_{{\leftmoon}}=5^{\circ}15$. In the same way,  $c_{\varepsilon}$ and $s_{\varepsilon}$ denote the cosine and sinus of the obliquity of the ecliptic $\varepsilon=23^{\circ}44$,  $i_{\star}=56^{\circ}06$ and $n$ denotes the mean motion of the test-particle, $n=\sqrt{\mu_{\Earth}/a^{3}}$.

 % Resonant Hamiltonian 	
 \begin{table}
 \caption{\label{Tab:HR}
 	Coefficients of the
 	resonant Hamiltonian $\mathcal{H}_{\R}$.}
 \center		
 \begin{tabular}{c|c|c}
 			\textrm{Terms}&
 			\textrm{Coefficients}&
 			\textrm{Values}\\
 			\hline
 			\hline
 			$X^2, Y^2$ & $c_{20},c_{02}$ & $\mp \frac{ 15 \mu_{\odot} (1 + c_{i_{\star}}) c_{\epsilon} s_{i_{\star}} s_\epsilon}{16 n r_{\odot}^3} \pm 	\frac{15 \mu_{{\leftmoon}} (1 + c_{i_{\star}}) c_{\epsilon} s_{i_{\star}} s_\epsilon (3 s_{i_{{\leftmoon}}}^2-2)}{32 n r_{{\leftmoon}}^3}$\\
 			
 			$X^2 Y^2$ & $c_{22}$ & $\frac{3 J_2 R_{\oplus}^2 ( 8 c_{i_\star}-5)}{32 a^4} - \frac{3 \mu_{\odot} ( 14 c_{i_\star}-9) ( 3 s_\epsilon^2-2)}{ 128 		a^2 n^2 r_{\odot}^3} + \frac{ 3 \mu_{{\leftmoon}} ( 14 c_{i_\star}-9 ) ( 3 s_{i_{{\leftmoon}}}^2-2) ( 3 s_\epsilon^2-2)}{ 256 a^2 n^2 				r_{{\leftmoon}}^3}$ \\
 			
 			$ X^4,Y^4$ & $c_{40},c_{04}$ & $\begin{array}{c} \frac{3 J_2 R_{\oplus}^2 ( 8 c_{i_\star}-5)}{64 a^4} - \frac{3 \mu_{{\leftmoon}} (3 s_{i_{{\leftmoon}}}^2 - 2) (26 - 26 c_{i_\star} - 60 c_{i_\star}^3 \pm 4 (17 c_{i_\star}-13 )  c_\epsilon s_{i_\star} s_\epsilon + 3 (13 c_{i_\star} + 30 c_{i_\star}^3-13) s_\epsilon^2)}{512 a^2 n^2 r_{{\leftmoon}}^3 ( c_{i_\star}-1)}  \\
 			+ \frac{ \mu_{\odot} (78 - 78 c_{i_\star} - 180 c_{i_\star}^3 \pm 12 ( 17 c_{i_\star}-13) c_\epsilon s_{i_\star} s_\epsilon + 9 (13 c_{i_\star} + 30 c_{i_\star}^3-13) s_\epsilon^2)}{ 256 a^2 n^2 r_{\odot}^3 ( c_{i_\star}-1 )}
 			\end{array}$ \\
 		\end{tabular}
 	\end{table}

 % CM Hamiltonian 	
 \begin{table}
 \caption{\label{Tab:HCM}
 Coefficients of the  centre-manifold Hamiltonian $\mathcal{H}_{\CM}$.}
 \center
 \begin{tabular}{c|c|c}
 			\textrm{Terms}&
 			\textrm{Coefficients}&
 			\textrm{Values}\\
 			\hline
 			\hline
 			$J_F$ & $ b_{10}$ & $ \frac{3 J_2 n R_{\oplus}^2 c_{i_\star}}{2 a^2} + \frac{ 3 \mu_{\odot} c_{i_\star} (2 - 3 s_\epsilon^2)}{
 				8 n r_{\odot}^3} + \frac{3 \mu_{{\leftmoon}} c_{i_\star} ( 3  s_{i_{{\leftmoon}}}^2 - 2) ( 3 s_\epsilon^2 - 2)}{16 n r_{{\leftmoon}}^3}$ \\
 			
 			$J_F^2$ & $ b_{20}$ & $-\frac{3 J_2 R_{\oplus}^2}{4 a^4}+ \frac{ 3 \mu_{\odot} ( 3 s_{i_{{\leftmoon}}}^2 - 2 ) }{16 a^2 n^2 r_{\odot}^3} - \frac{ 3 \mu_{{\leftmoon}} (3 s_{i_{{\leftmoon}}}^2 - 2) ( 3 s_\epsilon^2 - 2)}{32 a^2 n^2 r_{{\leftmoon}}^3}$ \\
 			
 			$\cos{u_F}$ &  $b_{01}$ & $ - \frac{3 a^2 \mu_{\odot} c_{i_\star} c_\epsilon s_{i_\star} s_\epsilon}{4 r_{\odot}^3} + \frac{ 3 a^2 \mu_{{\leftmoon}} c_{i_\star} s_{i_\star} c_\epsilon s_\epsilon (3 s_{i_{{\leftmoon}}}^2 - 2) }{8r_{{\leftmoon}}^3}$ \\
 			
 			$\cos{2u_F}$ & $ b_{02}$ &  $\frac{3 a^2 \mu_{\odot} (c_{i_\star}^2-1)s_\epsilon^2}{16 r_{\odot}^3} - \frac{3 a^2 \mu_{{\leftmoon}} (c_{i_\star}^2-1)(3 s_{i_{{\leftmoon}}}^2-2) s_\epsilon^2}{32 r_{{\leftmoon}}^3}$ \\
 			
 			$\cos(2 u_F + \Omega_{{\leftmoon}})$ & $ d_{21}$ & $ \frac{3 a^2 \mu_{{\leftmoon}} (c_{i_\star}^2 -1) c_{i_{{\leftmoon}}} (1 + c_\epsilon) s_{i_{{\leftmoon}}} s_\epsilon}{16 r_{{\leftmoon}}^3} $\\
 			
 			$\cos(u_F + \Omega_{{\leftmoon}})$ & $ d_{11}$ & $ -\frac{3 a^2 \mu_{{\leftmoon}} c_{i_\star} s_{i_\star} c_{i_{{\leftmoon}}} s_{i_{{\leftmoon}}} (1 + c_\epsilon -2 s_\epsilon^2)}{8 r_{{\leftmoon}}^3} $\\
 		\end{tabular}
 	\end{table}
 	
 % Coupling terms
 \begin{table}
 \caption{\label{Tab:HC}Coefficients of the  coupling terms Hamiltonian $\mathcal{H}_{\C}$.}
 \center
 \begin{tabular}{c|c|c}
 			\textrm{Terms}&
 			\textrm{Coefficients}&
 			\textrm{Values}\\
 			\hline
 			\hline		
 			$J_F X^2$, $J_F Y^2$ & $c_{120},c_{102}$ & $\begin{array}{c}  \frac{3 J_2 R_{\oplus}^2 ( 10 c_{i_\star} -1)}{8 a^4} + \frac{\mu_{\odot} ( 18 -54 c_{i_\star} \pm 30 (2 c_{i_\star} -1) c_\epsilon s_{i_\star} s_\epsilon + 27 (3 c_{i_\star}-1) s_\epsilon^2}{32 a^2 n^2 r_{\odot}^3 (c_{i_\star}-1)} \\ + \frac{3 \mu_{{\leftmoon}} (3 s_{i_{{\leftmoon}}}^2 -2)(18 c_{i_\star}-6 \pm 10 ( 1 - 2 c_{i_\star}) c_\epsilon s_{i_\star} s_\epsilon + 9 (1 - 3 c_{i_\star}) s_\epsilon^2)}{64 a^2 n^2 r_{{\leftmoon}}^3 (c_{i_\star}-1)} \end{array}$ \\
 		\end{tabular}
 	\end{table}
 \end{appendices}
 
 \section*{Acknowledgments} 
 J.D. is a postdoctoral researcher of the `Fonds de la Recherche Scientifique' - FNRS. 
 J.D. acknowledges the support of the ERC project $677793$ `Stable and Chaotic Motions in the
 Planetary Problem' leaded by Prof.\,Gabriella Pinzari. 
 I.G. acknowledges the support of the ERC project $679086$ COMPASS `Control for Orbit Manoeuvring
 through Perturbations for Application to Space Systems'.
 E.\,L. has been supported by the Marie Curie Initial Training Network Stardust-R, grant agreement Number $813644$ under the H2020 research and innovation program. C.\,E. was partially supported by EU-ITN Stardust-R and MIUR-PRIN 20178CJA2B `New Frontiers of Celestial Mechanics:
 theory and Applications'.

\bibliographystyle{apalike}
\bibliography{biblio}

\end{document}